\documentclass[%
 aip,
 amsmath,amssymb,
 reprint,%
]{revtex4-1}

\usepackage{graphicx}
\usepackage{dcolumn}
\usepackage{bm}

\usepackage[utf8]{inputenc}
\usepackage[T1]{fontenc}
\usepackage{mathptmx}
\usepackage{etoolbox}
\usepackage{mathtools}
\usepackage{xcolor}
\usepackage{epstopdf}
\usepackage{amsfonts,amssymb,amsmath,array}
\usepackage{amsbsy}
\usepackage{color}
\usepackage{hyperref} 
\usepackage{enumerate}
\usepackage{lipsum} 

\usepackage{bigints}
\usepackage{comment}
\usepackage{array}

\makeatletter
\def\@email#1#2{%
 \endgroup
 \patchcmd{\titleblock@produce}
  {\frontmatter@RRAPformat}
  {\frontmatter@RRAPformat{\produce@RRAP{*#1\href{mailto:#2}{#2}}}\frontmatter@RRAPformat}
  {}{}
}%

\newcommand{\lx} {\left}
\newcommand{\rx} {\right}

\newcommand{\ave}[1] {\lx\langle #1 \rx\rangle}

\makeatother

\begin{document}

\preprint{AIP/123-QED}

\title{Detecting time-irreversibility in multiscale systems: \\ correlation and response functions in the Lorenz96 model}
\author{Niccolò Cocciaglia}
\email{niccolo.cocciaglia@uniroma1.it}
\affiliation{Department of Physics, University of Rome Sapienza, P.le Aldo Moro 2, 00185, Rome, Italy}
\affiliation{INFN, Sezione di Roma ``Tor Vergata'', Via della Ricerca Scientifica 1, 00133, Rome, Italy}
\author{Dario Lucente}
\email{dario.lucente@unicampania.it}
\email{dariolucente32@gmail.com}
\affiliation{Department of Mathematics \& Physics, University of Campania “Luigi Vanvitelli”, 81100, Caserta, Italy}
\date{\today}


\begin{abstract}
Due to their relevance to geophysical systems, the investigation of multiscale systems through the lens of statistical mechanics has gained popularity in recent years. 
The aim of 
our work is the characterization of the nonequilibrium properties of the well-known two-scales Lorenz96 model, a dynamical system much used for testing ideas in geophysics, by 
studying either higher-order correlation functions or response to external perturbations of the energy. 
These tools in both equilibrium (inviscid) or non-equilibrium (viscous) systems
provide clear evidence of their suitability for detecting time-reversal symmetry breaking and for characterizing transport properties also in this class of models. 
In particular, we characterize how localized energy perturbations are transported between the different scales, highlighting that perturbations of synoptic variables greatly impact advective variables but perturbations of the latter have a practically negligible effect on synoptic scales. 
Finally, we show that responses of global observables to finite size perturbations strongly depend on the perturbation protocol. 
This prevents the physical understanding of the system from observations of the relaxation process alone, a fact often overlooked.
\end{abstract}

\maketitle


\begin{quotation}
Systems out of statistical equilibrium display a breaking of the time-reversal symmetry: the probability of a direct trajectory differs from that of its time-reversed counterpart. 
Due to the general difficulty to compute these probabilities, one often has to resort to alternative methods to detect time irreversibility. 
Asymmetric time correlations and non-diagonal response functions are suitable candidates, and they even provide additional insights to the system of interest. 
We will analyze the behaviour of both indicators for the two-scales Lorenz96 model, in its viscous (with external forcing and linear damping) and inviscid (with zero forcing and damping) versions, and emphasize the features emerging from the presence of slow and fast variables.
\end{quotation}


\section{Introduction}
The study of geophysical phenomena is notoriously challenging for both mathematicians and physicists. 
The inherent complexity can be attributed to the nontrivial interactions existing between the system components possessing different length and time scales, but also to the mere difficulty (even computational) of properly taking into account the enormous amount of variables needed to describe any relevant observable.
These difficulties could be overcome with a clever modelization of the phenomenon, having the twofold aim of simplifying the interactions appearing in the original equations and reduce the number of degrees of freedom. 
Naturally the procedure of discarding variables and interactions 
is rather delicate: the resulting equations still need to be a valid approximation of the original ones, and have to reproduce as faithfully as possible the features (either dynamical or statistical) of our system.

E. N. Lorenz clearly understood the advantages of simplified models \cite{Lorenz2005Designing} to gain some insight on complex phenomena, 
especially when dealing with geophysical flows. 
The ``Lorenz63 model" \cite{Lorenz63Deterministic}, a hyper-simplified modelization of thermal convection, had an enormous impact on chaos theory and fluid mechanics, and arguably represents the prototypical chaotic model for most computational studies. 

Less popular, yet with quite a considerable influence, is another model that became known as ``Lorenz96" \cite{Lorenz1996predictability}. 
The original aim was the study of atmospheric predictability, and the possibility to gain information about the spreading of initial small uncertainties by exploiting dynamical models attempting to reproduce atmospheric circulation. 

Lorenz presented two versions of the model. He first introduced a single-scale model where the dynamical variables, placed along a latitude circle, have asymmetric nonlinear couplings and are subject to constant forcing and linear damping. 
The combination of these three actions can roughly account for most geophysical phenomena taking place in the atmosphere. 
In the same work, Lorenz also proposed a two-scales version of the model, where the interacting variables have either slow or fast dynamics. 
Resolving the faster scales as well should make the model more adherent to real physics.
The wide influence of the Lorenz96 model is proved by the plethora of works which adopted it as the core of their studies or as a case study for applying new concepts. 
It was employed likewise in physics \cite{boffetta2000predictability, Lacorata2007fluctuation, Karimi2012extensive, Gallavotti2014equivalence, vankekem2018wave}, mathematics \cite{Orrell2003, Stappers2012, deLeeuw2018projected, kerin2020lorenz} and geosciences \cite{Basnarkov2012forecast, Sterk2012predictability, carlu2019lyapunov}, just to mention a few.

In this work we are interested in the characterization, in the two-scales Lorenz96 model, of temporal nonequilibrium properties as revealed by two effective indicators of the absence of statistical equilibrium: high-order time correlation functions \cite{Pomeau1982symetrie} and response functions \cite{kubo1966FDT, FDreport}. 
There are several reasons for choosing the Lorenz96 model for our purposes: the presence of nonlinear interactions between variables hinders the use of the solid theoretical framework built for linear systems, thereby making necessary the use of computer simulations. 
Since we have access to the explicit evolution equations, we can choose to study the `standard' model, namely a driven dissipative system, which is far from statistical equilibrium, or else the `reversible' model, where damping and external forcing are put to zero and the total energy is a global conserved quantity, so that the system is in statistical equilibrium in the microcanonical sense. 
In the following we will refer to the original model as ``viscous" or ``irreversible" Lorenz96, and to the model without forcing and damping as ``inviscid" or ``reversible" Lorenz96.
In this respect, it becomes especially interesting to analyze the differences between the two above-mentioned indicators when they refer to the former or the latter version of the model.

Non-symmetric third-order time correlation functions provide results that fit well with recent studies \cite{Lucente2023revealing, Cocciaglia2024nonequilibrium}, for what concerns their ability to discriminate between statistical equilibrium and nonequilibrium: 
correlation functions in the reversible model show very small 
deviations from zero, while the same functions in the irreversible model are characterized by a large short-time peak followed by relaxation to zero, thus the time dependence is more nontrivial. 
The second indicators, namely the response functions \cite{FDreport} of the `local' (related to a single d.o.f.) energies, reveal interesting dynamical properties. 
Specifically, the spreading of an initially-localized energy perturbation highlights the presence of statistical fluxes that behave as travelling waves. 
This happens only in the irreversible model, while in the reversible case the response functions display an asymptotic approach to equipartition, as expected from equilibrium statistical mechanics considerations.

The paper is structured as follows: in Sec.~\ref{sec:model} we introduce the model and delineate the main differences between the reversible and irreversible versions. In Sec~\ref{sec:unveil}, the statistical properties of the system are characterized through the lens of non-equilibrium statistical mechanics. In particular, Sec.~\ref{subsec:higher-order} deals with higher-order correlation functions while Sec.~\ref{subsec:responses} examines the responses of the system to external perturbations. Finally, in Sec.~\ref{sec:conclusion} we draw the main conclusions and present possible perspectives.


\section{Model}\label{sec:model}
Among the many variants present in the literature, we decided to focus on the original multiscale version proposed by Lorenz~\cite{Lorenz1996predictability}.
The model features the evolution of $N$ slow variables $\{X_n\}$ ($n=1,\, \dots\, ,N$) interacting with a set of $N \cdot K$ fast variables $\{y_{n,k}\}$ ($k=1,\, \dots\, , K)$. 
More precisely, the $\{X_n\}$ 
represent the values of generic coarse-grained synoptic observables over $N$ sectors of a latitude circle, while $\{y_{n,k}\}$ denote the values taken by a convective field within $K$ subsectors of the $n-$th latitude sector. The system thus evolves in a $D-$dimensional space (with $D=(K+1)N$) according to the following laws:
\begin{align}
    \dot{X}_n &= X_{n-1}\lx(X_{n+1}-X_{n-2}\rx)-\frac{hc}{b}\sum_{k=1}^Ky_{n,k}-\nu_1X_n+F\nonumber\\
    \dot{y}_{n,k} &= cb\, y_{n,k+1}\lx(y_{n,k-1}-y_{n,k+2}\rx)+\frac{hc}{b}X_n-c\,\nu_2\,y_{n,k}
    \label{eq:model}
\end{align}
where the parameter $c$ controls the relative timescale between fast and slow variables, $b$ the relative amplitude, while $h$ represents the coupling strength. 
In the following we consider $c=5$, $b=10$, $h=1$, $N=30$ and $K=5$. 
The viscous case has $F=10$ and $\nu_1 = \nu_2 = 1$, and such values place the system in the chaotic regime \cite{frank2014standing}, whereas in the inviscid case we put $F=\nu_1=\nu_2=0$.
The equations for the slow and fast variables have a similar structure with quadratic non-linearities and viscous dissipation but differ with respect to the couplings: each $X_n$ feels the average effect of the fast variables $\{y_{n,k}\}$ in the same latitude sector $n$, while each $y_{n,k}$ perceives the synoptic field $X_n$ as a forcing term. 
A further difference concerns the forcing $F$ that acts only at slow scales.
Note that Eqs.~\eqref{eq:model} are, in form, similar to a class of simplified turbulent systems called shell models \cite{bohr1998dynamical,biferale2003shell,ditlevsen2010turbulence}, although in the former the subscripts $n$ and $k$ indicate a sort of spatial structure, while in the latter the `shell variables' reproduce spectral velocity components.
In any case, in both models non-linear terms are introduced to mimic advection, while an external forcing and a linear dissipative term are necessary for reproducing the out-of-equilibrium stationary state resulting from their balance.
In this sense, also the Lorenz96 model belongs to the class of driven-dissipative systems, widely employed for describing geophysical phenomena.
The advective 
character of the non-linearities appears evident considering the time-evolution of the total energy $E_{tot}$, that is
\begin{align}
&E_{tot}=\frac{1}{2}\sum_{n=1}^N(X_n^2+\sum_{k=1}^K y_{n,k}^2)\,,\nonumber\\ 
&\frac{{\rm d}E_{tot}}{{\rm d}t}=\sum_{n=1}^N\left(X_n F- \nu_1 X_n^2-c\,\nu_2\sum_{k=1}^K y_{n,k}^2\right)\,.
\label{eq:Energy_tot-Variation}
\end{align}
As can be seen from the above expressions, the energy varies due to the forcing and dissipation terms only, and it is constant when $F$, $\nu_1$ and $\nu_2$ are set to zero. 
\begin{figure*}[ht]
\centering
\includegraphics[width=0.49\textwidth]{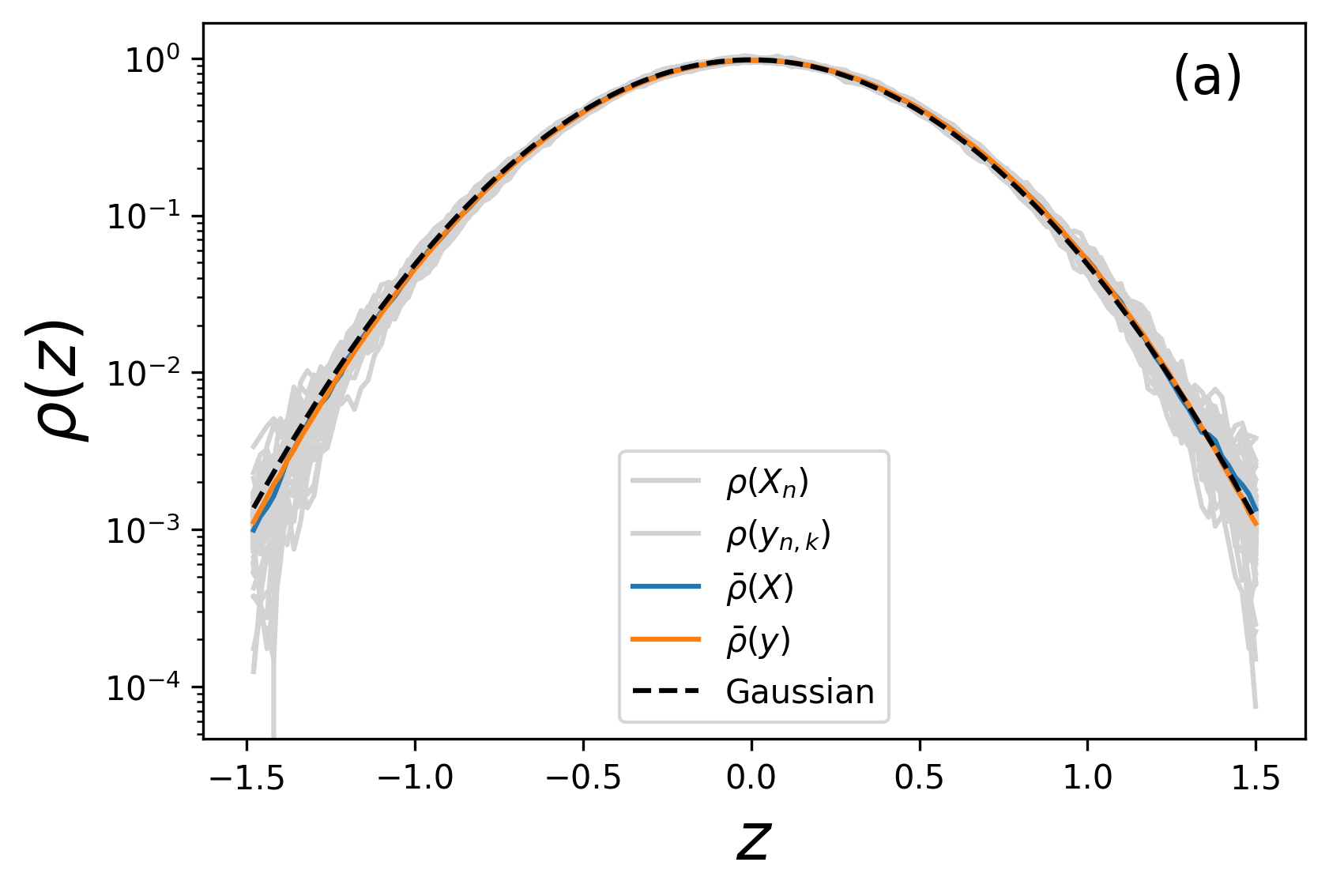}
\includegraphics[width=0.485\textwidth]{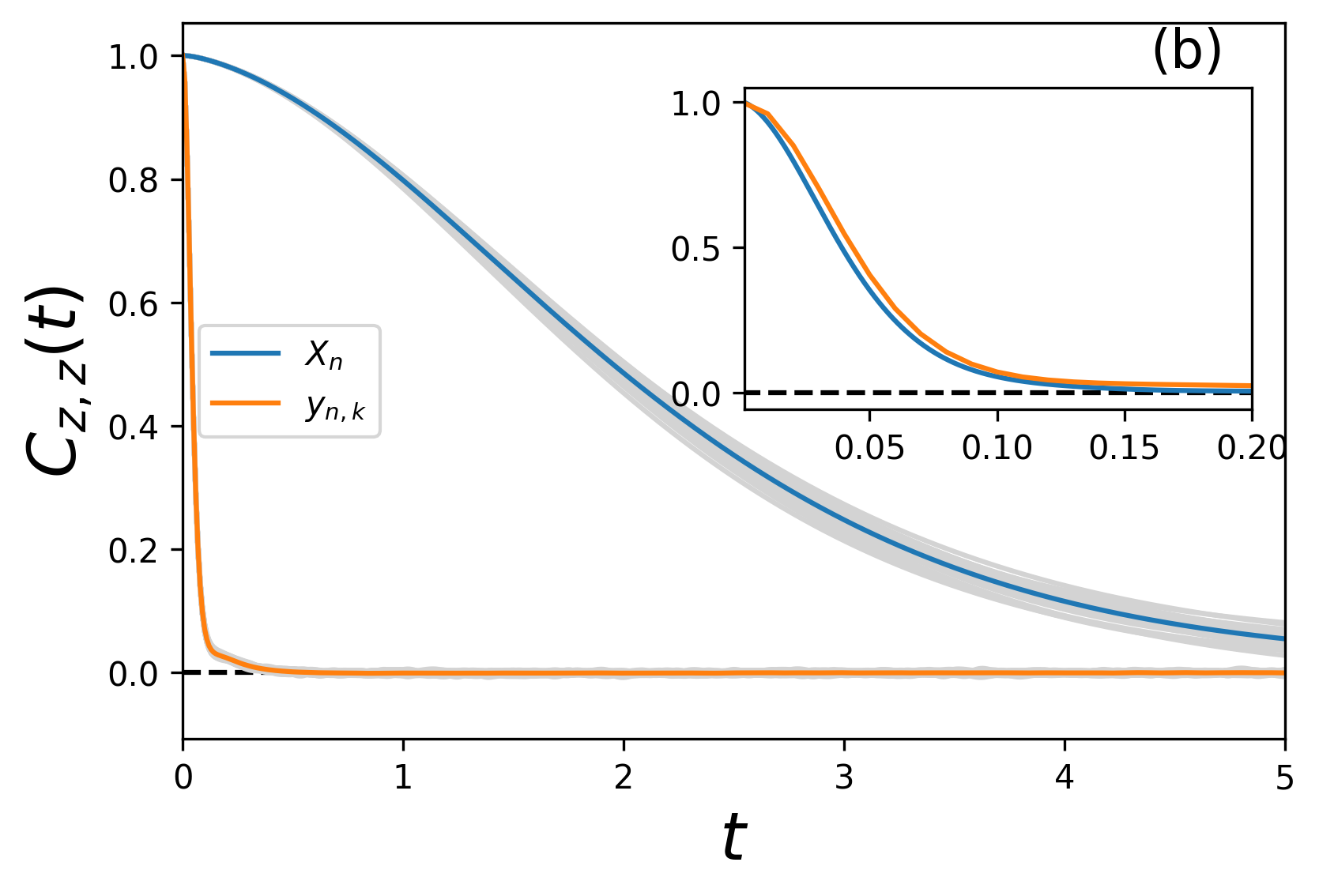}
\vspace{-3.5mm} 
\caption{Reversible system. (a): Probability density functions $\rho(z)$ where $z$ denotes both fast and slow variables. Light grey lines indicate pdf's obtained from single-variable trajectories, blue and orange lines denote the average pdf $\bar{\rho}(z)=\frac{1}{M}\sum_n \rho(z_n)$ for $z=X,y$ and $M = N, N\cdot K$, respectively. Gaussian distribution is represented by a black dashed line. (b): Time autocorrelation functions $C_{z,z}(t)$. Light grey curves are individual autocorrelations, the blue one is the average autocorrelation computed on all slow variables, and the orange curve the same for fast variables. The inset shows the collapse obtained when the time axis of the blue curve is rescaled by a factor $bc$.}
 \label{fig:Distribution_Correlation_Equilibrium}

\vspace{4.5mm} 
\includegraphics[width=0.48\textwidth]{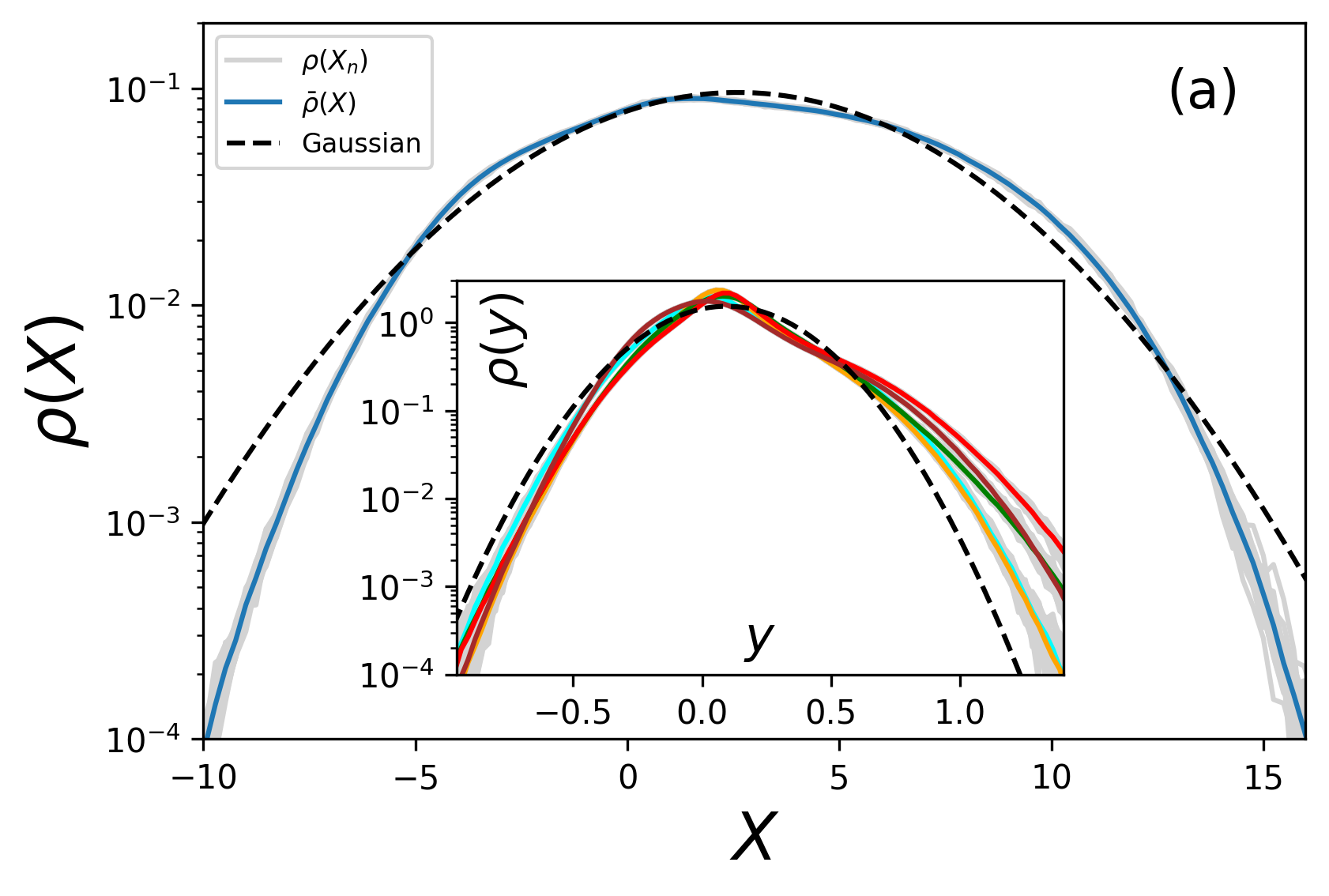}\includegraphics[width=0.49\textwidth]{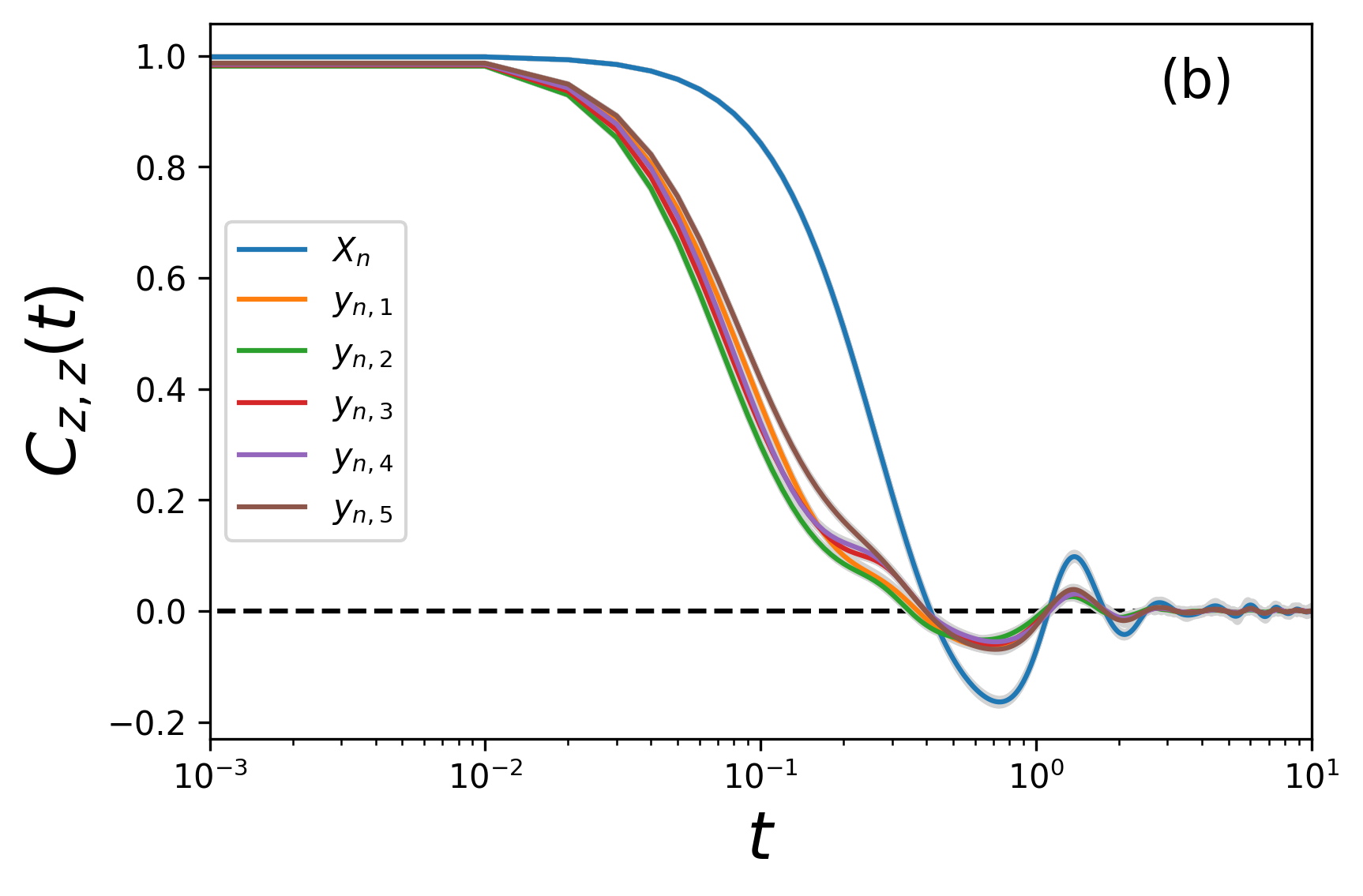}
\vspace{-3.5mm} 
\caption{Irreversible system. (a): Probability density functions $\rho(z)$ for $z=X$ (main plot) and $z=y$ (inset). Light grey lines indicate pdf's obtained from single-variable trajectories, the blue line denotes the average pdf $\bar{\rho}(X)=\frac{1}{M}\sum_n \rho(X_n)$ while different colors in the inset represent the same quantity for different indices of the fast variable ($z=y_{n,k}$ for $k=1,\, \dots\, ,K$). Gaussian distributions are represented by black dashed lines. (b): Semilog plot of the time autocorrelation functions $C_{z,z}(t)$, for single-variables (grey) or averaged over indices $n$, for both $z=X_n$ (blue) and $z=y_{n,k}$ (each color representing a different index $k=1,\, \dots\, ,K$).}
 \label{fig:Distribution_Correlation_Non-Equilibrium}
\end{figure*}

In the inviscid version, despite the non-Hamiltonian structure of the problem, the system evolves on a $(D-1)$-dimensional surface with constant energy $E_{tot}$ determined by the initial conditions. 
Hence,
all statistical properties of the system are described by the microcanonical ensemble. 
Furthermore, since $E_{tot}$ is quadratic in $X$ and $y$, the equipartition theorem holds, i.e. 
\begin{equation}
\ave{X_n^2}=\ave{y_{n,k}^2}=\frac{E_{tot}}{N(K+1)}=\frac{E_{tot}}{D}\, .
\end{equation}
The dynamics is still chaotic like in the viscous forced case, and a computation of $\lambda_1$, the largest Lyapunov exponent, gave the value $\lambda_1 \simeq 3.7$.
The system is homogeneous and each variable is distributed according to a normal distribution with variance $E_{tot}/D$, as can be seen in Fig.~\ref{fig:Distribution_Correlation_Equilibrium}a. 
Yet the equivalence of the variables holds only with respect to static statistical properties, such as the average energy. 
From a dynamical point of view, on the other hand, the variables $X_n$ and $y_{n,k}$ evolve on different time scales, as clearly highlighted by Fig.~\ref{fig:Distribution_Correlation_Equilibrium}b showing the 
normalized time autocorrelation functions
 \begin{equation}
     C_{z,z}(t) \equiv \frac{\langle z(t_0) z(t_0+t) \rangle - \langle z(t_0) \rangle^2}{\langle z^2(t_0) \rangle - \langle z(t_0) \rangle^2}\ .
 \end{equation}
Here angular brackets 
denote ensemble average, replaced by temporal average with respect to $t_0$ invoking ergodicity and assuming statistical stationarity, 
while $z=\{X_n, y_{n,k}\}$ indicates a generic variable of the system.
Since in the inviscid case the system is subject only to advection, and the advective terms in the equations for the $y$'s are $bc$ times larger than those of the $X$'s, the ratio between the two time scales should be exactly $bc$. 
This is confirmed in the inset of Fig.~\ref{fig:Distribution_Correlation_Equilibrium}b, where the slow-variable autocorrelation is plotted as a function of the rescaled time $t\, /\, bc$, showing a nice collapse onto the fast-variable curve.

The behavior of the viscous system is rather different from the inviscid case: the total energy is no longer a constant of motion and, after a transient, it fluctuates around a stationary value that is determined by the balance between forcing and dissipation. 
This causes the system to behave irreversibly and the statistical properties are no longer described by an equilibrium statistical ensemble. 
In this situation the equipartition theorem does not hold, and fast and slow variables cease to be equivalent even with regard to static properties. 
Interestingly, although the system is still homogeneous with respect to the $N$ sectors of the latitude circle (i.e. all $X_n$ are equivalent), it becomes inhomogeneous within the $NK$ subsectors so that the fast variables $y_{n,k}$ and $y_{n',k'}$ are statistically equivalent only for $k=k'$. 
All these considerations are illustrated in Fig.~\ref{fig:Distribution_Correlation_Non-Equilibrium}a, which shows the probability distributions for both $X_n$ (main plot) and $y_{n,k}$ (inset). 
First, it emerges that both variables are skewed and clearly non-Gaussian \cite{Lacorata2007fluctuation}. 
\begin{figure*}[t]
    \centering
    \includegraphics[width=\textwidth]{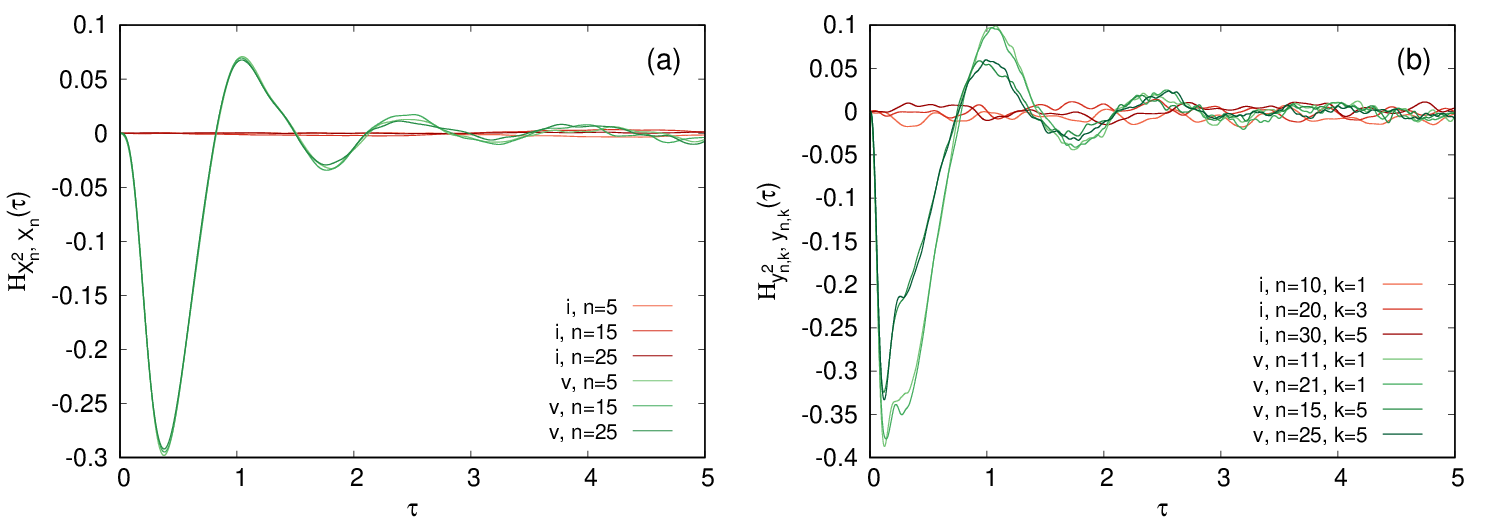}
    \vspace{-6mm}
    \caption{Asymmetric time correlation functions in the inviscid (red palette, labelled with `i') and viscous (green palette, labelled with `v') Lorenz96 model for different $n$'s. (a) Correlations of slow variables,  Eq.~\eqref{eq:pomeau_z} with $z=X_n$. (b) Correlations of fast variables, Eq.~\eqref{eq:pomeau_z} with $z=y_{n,k}$. Non-overlap between leftmost ($k=1$) and rightmost ($k=K$) fast variables, for arbitrary $n$, is evident. 
    }
    \label{fig:pomx2x}
\end{figure*}
Furthermore, comparing the horizontal axes, it can be seen that the $y$ variables have a standard deviation approximately $b$ times smaller than the one of the $X$ variables, as imposed by construction in the model.  
Regarding the dynamical behavior, the autocorrelation functions (Fig.~\ref{fig:Distribution_Correlation_Non-Equilibrium}b) of both $X$ and $y$ display, at later stages, oscillations with the same frequency and phase.
However, the autocorrelation functions of the $y$'s decay roughly $c$ times faster than the correlations of the $X$'s. 
The short-time behaviour is due to the fact that the characteristic dissipation times of $X$ and $y$ differ by exactly a factor of $c$, but at larger times the slower timescale, that of the $X$'s, will drive the fast variables and induce low-frequency oscillations in them.

As we will see in the next section, the properties of the model mentioned above will play a crucial role in interpreting the behavior of the indicators which are typically considered to reveal time irreversibility.

\section{Unveiling non-equilibrium}\label{sec:unveil}

In this Section, we focus on the problem of detecting time-reversal symmetry breaking in the two-scales Lorenz96 model. 
There are different indicators that allow us to quantity the degree of non-equilibrium in a generic physical system and the choice of one indicator over another often depends on the available information. 
From a mathematical point of view, the most suitable measure is the entropy production~\cite{lebowitz1999microscopic,lebowitz1999gallavotti} which compares the probabilities of forward-in-time and backward-in-time trajectories. 
While conceptually advantageous due to its reference frame independence, this quantity is typically not experimentally accessible~\cite{Lucente2023poisson}. 
To detect temporal irreversibility without having to face technical limitations related to entropy production, it is possible to consider suitable asymmetric time correlation functions \cite{Pomeau1982symetrie}: the effectiveness and versatility of such method is proved by its widespread use in different fields, like stochastic processes \cite{Lucente2023poisson}, turbulence \cite{Josserand2017, Cocciaglia2024nonequilibrium} and granular materials \cite{Lucente2023revealing}, but also in econometrics \cite{Ramsey1996business}. Another important tool of non-equilibrium statistical mechanics is represented by response theory~\cite{kubo2012statistical} due to its connection to fluctuation-dissipation relations (FDRs). As we will see in subsection~\ref{subsec:responses}, response functions not only provide invaluable information regarding statistical fluxes, but it is also possible to determine whether or not the underlying system is in thermal equilibrium by inspecting their asymptotic behaviour. 



\subsection{Higher-order correlations}\label{subsec:higher-order}

Correlation functions are the simplest objects able to detect temporal irreversibility from experimental records. 
To be more formal, let us consider two statistically stationary observables (or time signals) $A(t)$ and $B(t)$ with an associated time correlation function $\tilde{C}_{A,B}(\tau) \equiv \langle A(t) B(t+\tau) \rangle$ (we added the tilde to distinguish it from the normalized one introduced before).  
Since equilibrium systems are invariant under time-reversal, the equivalence $\tilde{C}_{A,B}(\tau) = \tilde{C}_{A,B}(-\tau)=\tilde{C}_{B,A}(\tau)$ holds for any pairs of observables $A$ and $B$. On the contrary, if there exist two observables $A$ and $B$ that are not invariant under time reversal operation $t \rightarrow -t$ and such that $\tilde{C}_{A,B}(\tau) \neq \tilde{C}_{B,A}(\tau)$, then the system is necessarily out-of-equilibrium~\cite{Pomeau1982symetrie,Lucente2023revealing,Lucente2023poisson,Cocciaglia2024nonequilibrium}.
The idea is then to measure deviations from equilibrium (or time-reversibility) by defining an asymmmetric time correlation function as:
\begin{equation}
    H_{A,B}(\tau) = \tilde{C}_{A,B}(\tau) - \tilde{C}_{B,A}(\tau)\, .
    \label{eq:H}
\end{equation}

Our aim here is to study the difference observed in the function $H$ when measured both in the inviscid model with $F=0$ and in the viscous one with $F>0$. 
The former is a conservative system, thereby its average properties are describable within the formalism of equilibrium statistical mechanics. 
The latter, a forced-dissipative system, deviates from equilibrium due to the presence of energy sources and sinks. 
The type of asymmetric time correlation functions worth investigating are, in non-dimensional form:
\begin{equation}
    H_{z^2,z^{}}(\tau) = \frac{\tilde{C}_{z^2,z^{}}(\tau) - \tilde{C}_{z^{},z^2}(\tau)}{\langle z^2(t) \rangle^{3/2}}\ ,
    \label{eq:pomeau_z}
\end{equation}
where again $z=\{X_n, y_{n,k}\}$ is a generic variable.
They represent the lowest-order type of non-symmetric time correlations that are local in terms of $X_n(t)$ or $y_{n,k}(t)$. 
The denominators need to be even moments of the variables, because odd-order ones vanish in the inviscid case.

The difference between the two 
systems clearly emerges in Fig.~\ref{fig:pomx2x}, representing in panel (a) the functions \eqref{eq:pomeau_z}, for $z=X_n$, for both forced and non-forced models and in panel (b) the functions \eqref{eq:pomeau_z}, for $z=y_{n,k}$, for the same cases. 
The large fluctuations experienced by the forced dissipative model are a clear sign of deviation from equilibrium, and the peculiar damped oscillations 
can be understood as a superposition of complex exponentials emerging from discrete Fourier transformation of the $X$'s and $y$'s, a procedure often used in the Lorenz96 model \cite{Lorenz1998optimal, MajdaWang2006}. 
As expected, the symmetry of slow variables under cyclic translation of the indices is reflected in the almost-perfect superposition of the correlations in the dissipative case. For the fast variables there is cyclic symmetry as well but only for translations modulo $K$, as emerges in Fig.~\ref{fig:pomx2x}b.

We are not able to attribute a clear physical meaning to the behaviour at short time, showing a consistent decrease of the functions $H$.
Let us also notice that the differences in typical magnitudes and timescales between $X$'s and $y$'s are lost in these time correlations: the peak amplitudes and oscillation frequencies are almost identical in the viscous case. 
Similarity in the peaks amplitudes is a simple consequence of the normalization, while the coincidence of timescales after a few decorrelation times was a trend already found in Fig.~\ref{fig:Distribution_Correlation_Non-Equilibrium}b for $C_{z,z}(\tau)$, and it can be explained in the same way.

We computed also asymmetric correlations of the sixth-order of the kind $H_{X^4_n,X^2_n}(\tau)$ and $H_{y^4_{n,k},y^2_{n,k}}(\tau)$, representing third-order time correlations of the local energies. 
The qualitative behaviour is 
rather similar to Fig.~\ref{fig:pomx2x}, despite the signals being more noisy since larger statistics are needed for computing higher-order moments (not shown). 
This is interesting because this form of the asymmetric correlations are formally equivalent to third-order moments of local-energy differences in time \cite{Cocciaglia2024nonequilibrium}, therefore the initial decreasing of $H_{z^4,z^2}(\tau)$ would represent a short-time energy loss for variable $z$. 
Yet this implies that both fast and slow variables lose energy at small times, while the total energy remains stationary. 
This apparent paradox can be solved by considering that, as suggested by further simulations, energy transfers actually involve the energy of a slow variable $X_r$ and the \textit{sum of} the energies of the corresponding slow ones $y_{r,k}$.
This is a matter still under investigation, so we will not delve into more details here.


\subsection{Responses to external perturbations}\label{subsec:responses}

In the context of nonequilibrium statistical mechanics (or nonlinear dynamics as well) little attention has been paid to a powerful tool, able to determine causal interactions \cite{Baldovin2020understanding} and transition asymmetries \cite{Sarra2021response}: 
\textit{non-diagonal response functions}. 
While the response function formalism found wide applicability also in far-from-equilibrium situations, especially in the form of Fluctuation-Dissipation Relations (FDRs) \cite{Ruelle1998general, Lucarini2008response, FDreport}, most theoretical and numerical interest was directed towards \textit{diagonal response functions} \cite{FDreport} for their connection with time autocorrelation functions and hence with decorrelation times of chosen variables.
There exist previous studies on response functions in the Lorenz96 model, with different aims: FDRs and stochastic modeling \cite{Lacorata2007fluctuation}, application of Ruelle linear response theory to climate \cite{Lucarini2011statistical}, predictions and causality \cite{Lucarini2018revising}.

Our aim here is to 
investigate the relaxation behaviour of a perturbation acting locally on a single variable $X_n$ or $y_{n,k}$. 
This relaxation will propagate to all the variables interacting directly or indirectly with the perturbed one, but the way in which the redistribution occurs provides useful information about the statistical properties of the system. 
As will be clear shortly, both the long-time behaviour of the response functions and the transients leading to it are able to discriminate between equilibrium and nonequilibrium, and highlight statistical fluxes where present. 
Of course the effectiveness of response functions for this type of study depends on the choice of the right observables.

Let us briefly introduce the type of response functions which will be the object of this study.
Broadly speaking, one is interested in how a perturbation $\delta A_m$ on the $m$-th component of a vector-valued dynamical observable $\boldsymbol{A}(t)$ affects on average one component of another (or the same) observable $\boldsymbol{B}(t)$, e.g. $B_n(t)$. 
An impulsive perturbation generates a perturbed trajectory of the system which, in chaotic systems, will separate exponentially from the unperturbed one as time evolves. 
Assuming the impulsive perturbation at $t=0$, its average effect on $B_n$ can be quantified with a response function defined as:
\begin{equation}
    R_{A_m\hspace{.1em} ;\hspace{.1em} B_n}(t) = \frac{\overline{\delta B_n}(t)}{\delta A_m(0)}\ ,
    \label{eq:general_rf}
\end{equation}
where the overbar indicates an average over realizations of the same initial perturbation and $\delta B_n (t) = B'_n(t) - B_n(t)$, where $B'_n(t)$ is the value computed along the perturbed trajectory at a time $t$ after the perturbation. 
\begin{figure}[t]
    \centering
    \includegraphics[width=0.5\textwidth]{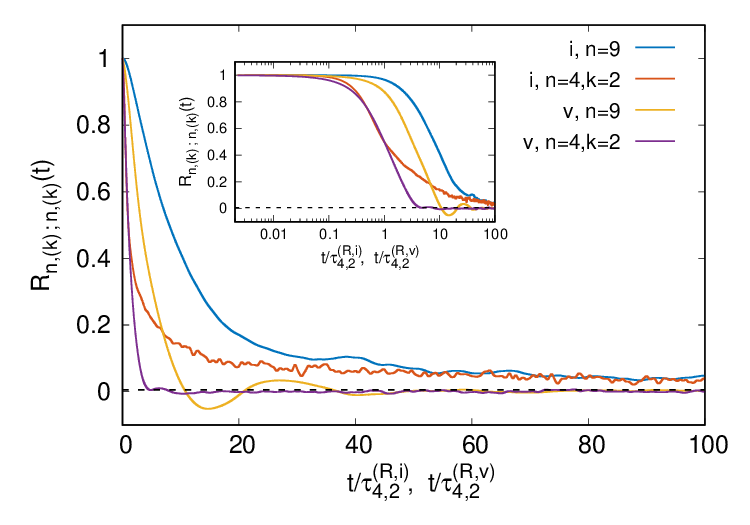}
    \vspace{-9mm}     
    \caption{Diagonal response functions for the energies of fast and slow variables, in both inviscid and viscous systems. Black dashed line: predicted equilibrium asymptote $\sim [N(K+1)]^{-1}=1/180$. Time rescaled with the relaxation times of the fast variables in inviscid ($\tau^{(R,i)}_{4,2}$) and viscous $(\tau^{(R,v)}_{4,2})$ cases. Their ratio is: $\tau^{(R,i)}_{4,2}/\tau^{(R,v)}_{4,2} \simeq 3.7$. Inset: same as main but in semilog scale. Here and in all following plots the average is over $10^5$ realizations.}
    \label{fig:diag_resp}
\end{figure}
\begin{figure}[t]
    \centering
    \includegraphics[width=0.5\textwidth]{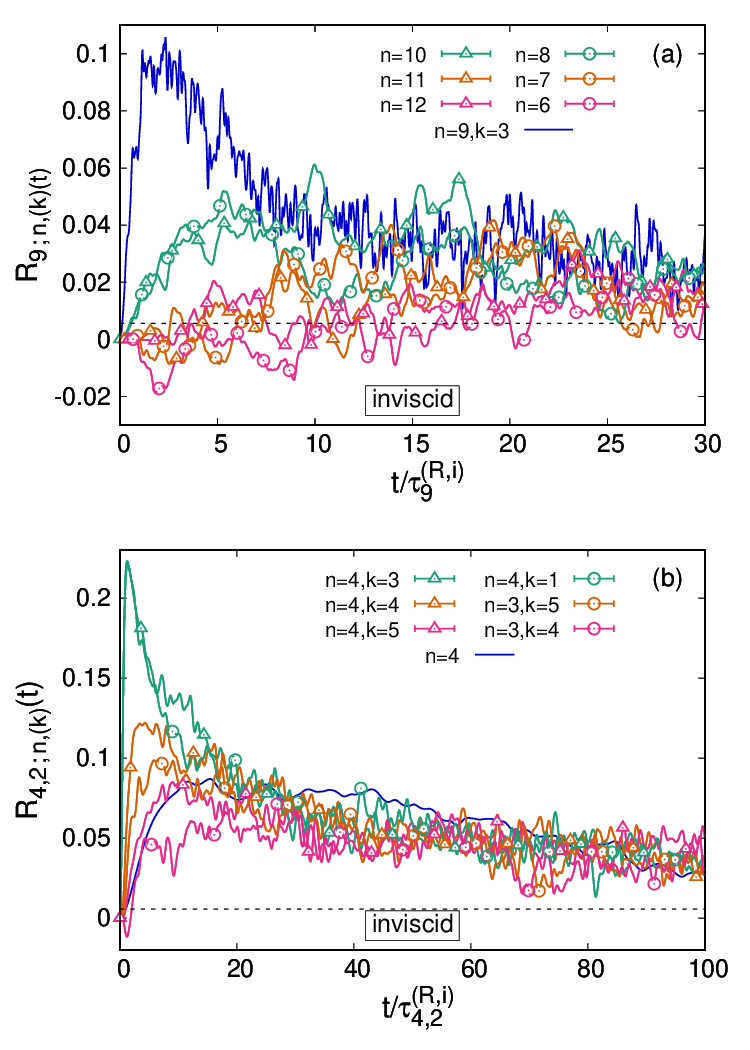}
    \vspace{-9mm}
    \caption{Nondiagonal response functions for the energies in the inviscid system.  Responses with same distance from the perturbed variable share the same color coding. (a) Perturbation on the energy of a slow variable. (b) Perturbation on the energy of a fast variable. In both panels, time is rescaled with the relaxation time of the perturbed variable. Triangles: intra-scale responses measured eastward. Circles: intra-scale responses measured westward. Blue solid line: inter-scale response of the corresponding (same $n$) variable. Black dashed line: same asymptote as Fig.~\ref{fig:diag_resp}.}
    \label{fig:nondiag_resp_eq}
\end{figure}
A reasonable choice could be to consider the local energies,
\begin{equation}
    E_{X_n} = X_n^2/2 \ \qquad \ E_{y_{n,k}} = y_{n,k}^2/2\ ,
    \label{energies}
\end{equation}
as the perturbed and measured observables of the response functions. 
The 
reason for such a choice is that it is physically easier to imagine a perturbation that increments or decreases the energy of a system, while a perturbation displacing directly the $X_n$'s or $y_{n,k}$'s could give problematic interpretations. 
In order to ease the notation, we will indicate as subscripts of the response functions only the indices of the variables, namely: $R_{m\hspace{.1em};\hspace{.1em} n}(t) \equiv R_{E_{X_m}\hspace{.1em};\hspace{.1em} E_{X_n}}(t)$ and $R_{m,j\hspace{.1em};\hspace{.1em} n,k}(t) \equiv R_{E_{y_{m,j}}\hspace{.1em};\hspace{.1em} E_{y_{n,k}}}(t)$. 
The 
indices before of after the semicolon makes clear if a fast or a slow variable is involved.
Thus, the focus will be on functions like:
\begin{align}
    R_{m\hspace{.1em};\hspace{.1em} n}(t) &= \frac{\overline{\delta E_{X_n}}(t)}{\delta E_{X_m}} \label{eq:resp_XX} \\ 
    R_{m,j\hspace{.1em};\hspace{.1em} n,k}(t) &= \frac{\overline{\delta E_{y_{n,k}}}(t)}{\delta E_{y_{m,j}}}\ ,
    \label{eq:resp_yy}
\end{align}
measuring intra-scale properties, or like:
\begin{align}
    R_{m\hspace{.1em};\hspace{.1em} n,k}(t) &= \frac{\overline{\delta E_{y_{n,k}}}(t)}{\delta E_{X_m}} \label{eq:resp_Xy} \\[2pt]
    R_{m,j\hspace{.1em};\hspace{.1em} n}(t) &= \frac{\overline{\delta E_{X_{n}}}(t)}{\delta E_{y_{m,j}}}\ ,
    \label{eq:resp_yX}
\end{align}
measuring inter-scale properties (the $t=0$ in the denominators has been omitted). 
\begin{figure*}[t]
    \centering
    \includegraphics[width=.95\textwidth]{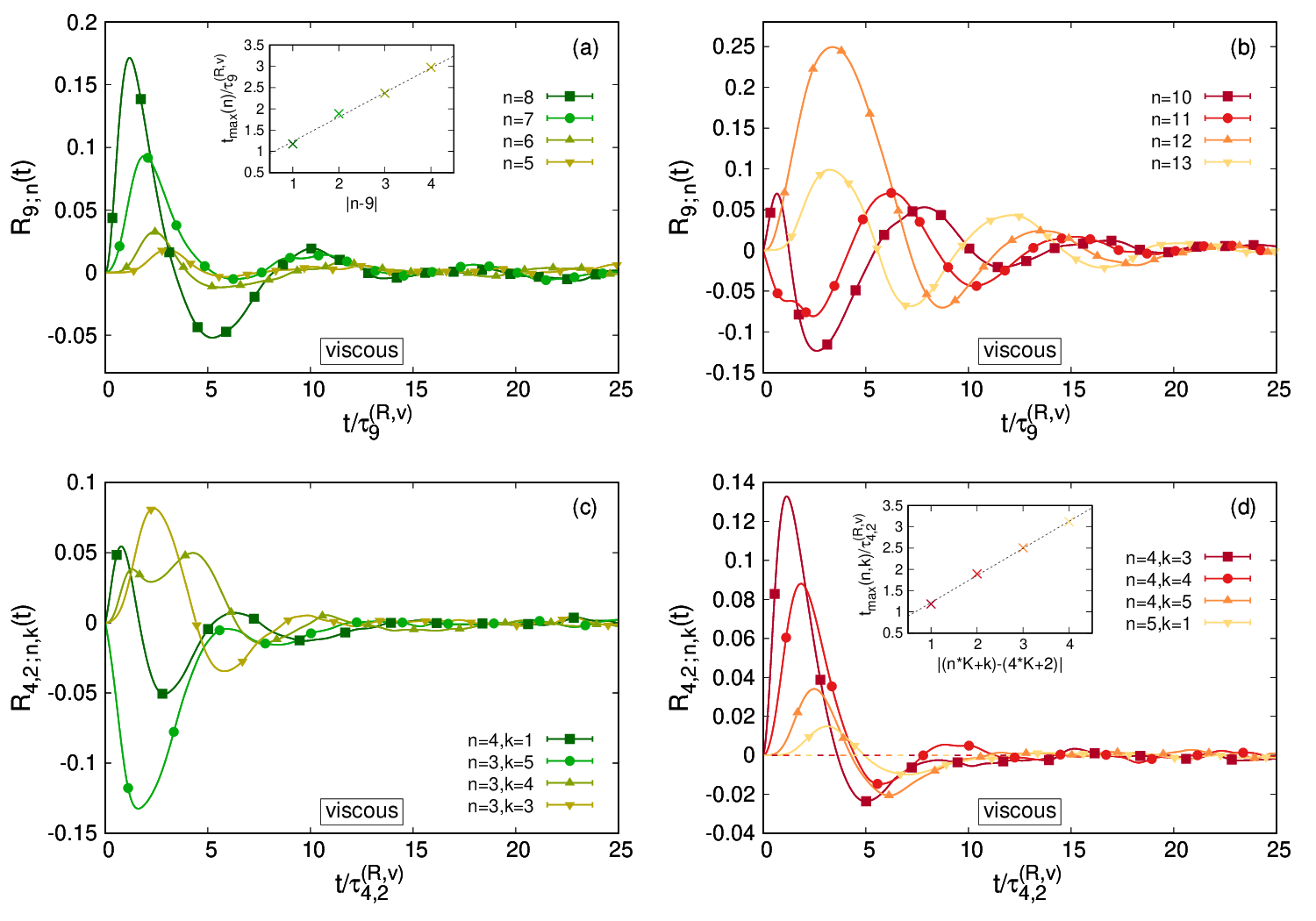}
    \vspace{-5mm}
    \caption{Nondiagonal energy response functions in the viscous system. Westward and eastward responses are represented separately. (a) Westward response functions for the perturbation on a slow variable. (b) Eastward response functions for the perturbation on the same slow variable. (c) Westward response functions for the perturbation on a fast variable. (d) Eastward response functions for the perturbation on the same fast variable.
    Time axis is rescaled like in Fig.~\ref{fig:nondiag_resp_eq}. Insets in panel (a) and (d) show the (rescaled) times at which the maxima are reached, as a function of the distance from the perturbation. The dashed lines are linear best fits.}
    \label{fig:nondiag_resp_noneq}
\end{figure*}
The (positive-signed) perturbations are performed in the following way:
\begin{equation}
    X'_m = \text{sgn}(X_m)\sqrt{X_m^2 + 2\ \delta E_{X_m}}
    \label{eq:perturbation}
\end{equation}
and similarly for $y_{m,j}$. Here $\delta E_{X_m}$ is the amplitude of the perturbation, chosen to be of the order of typical fluctuations of that energy \cite{Rfinpert}, namely equal to the standard deviation of $E_{X_m}$. 
This procedure ensures that the perturbation is independent of $X_m$ and it always increases its energy.

Diagonal response functions provide quantitative information about the relaxation time of the initial perturbation: 
usually one can define the relaxation time $\tau^{(R)}_{m}$ (in the same compressed notation) as the time a diagonal response function $R_{m\hspace{.1em};\hspace{.1em} m}(t)$ takes to cross a chosen threshold, e.g. $1/2$. 
In Fig.~\ref{fig:diag_resp} the four different kinds of self response functions are shown, with time measured in terms of the relaxation times of the perturbation on $E_{y_{4,2}}$ in both inviscid and viscous systems, named respectively $\tau^{(R,i)}_{4,2}$ and $\tau^{(R,v)}_{4,2}$. 
The relaxation is slower in the inviscid system: not only $\tau^{(R,i)}_{4,2}$ is almost 4 times bigger than $\tau^{(R,v)}_{4,2}$, 
but comparing the relaxation times between slow and fast variables of same-type systems we find $\tau^{(R,i)}_{9}/\tau^{(R,i)}_{4,2} \simeq 8.0$, whereas $\tau^{(R,v)}_{9}/\tau^{(R,v)}_{4,2} \simeq 3.2$. (see blue and yellow curves in Fig.~\ref{fig:diag_resp}). 
The different relaxation times are appreciated better in the inset, where a logarithmic time axis is employed.
The comparison between the characteristic time scales of temporal autocorrelation functions and diagonal responses deserves some discussion. 
{\color{black}
In the viscous case, both the ratio of slow-to-fast-variables decorrelation times and the energy relaxation time of a slow variable divided by that of a fast variable are close to the value of $c$.
In the inviscid system, instead, the ratio between the time scales of the energy self-RFs is approximately $10$, while for the variables autocorrelations a ratio of about $bc=50$ was estimated in Paragraph~\ref{sec:model} (see Fig.~\ref{fig:Distribution_Correlation_Equilibrium} and related discussion). 
Even though FDRs cannot be invoked, since neither we are in the linear regime nor we are looking at time correlations and responses on the same observables (namely, correlations of $z$ and response functions of $z^2$), it is nonetheless interesting to note how differently multiscale systems behave with respect to time-dependent statistical properties when one switches from a conservative to a dissipative dynamics.}

There is a clear difference in the asymptotic behaviour of equilibrium and non-equilibrium response functions. While the viscous responses relax to zero, meaning that the perturbation will eventually die out and the perturbed energy will recover its unperturbed value, the inviscid ones approach a common positive asymptote. 
Indeed, equilibrium statistical mechanics tells us that a perturbation on a conservative system will carry it away from the hypersurface in phase space where the dynamics was constrained to a new ``perturbed'' hypersurface. 
Assuming that the asymptotic state is a new equipartition configuration, then the long time value of all inviscid response functions reads: $R_{m,(j)\hspace{.1em};\hspace{.1em} n,(k)}(t) \xrightarrow{t\rightarrow\infty}[N(K+1)]^{-1}$. 
Due to the slow dynamics the establishment of a new equipartition state takes a very long time, so even our longest computations weren't able to get close to the expected asymptotic value.
Also inviscid nondiagonal responses tend to reach the same positive 
asymptotic value: in Fig.~\ref{fig:nondiag_resp_eq} the average spreading of an energy perturbation, performed respectively on a slow (Fig.~\ref{fig:nondiag_resp_eq}a) and a fast (Fig.~\ref{fig:nondiag_resp_eq}b) variable, are represented. 
Also instances of inter-scale response functions are included as solid blue lines.

We will now focus on the response functions in the viscous system, with a constant forcing on each of the $X_n$'s and linear damping on all variables. 
It is known \cite{Lorenz1998optimal, MajdaWang2006} that, when the viscous system is in the chaotic regime, localized fluctuations around the steady state propagate like waves with phase velocity and group velocity having opposite directions. 
This feature is a consequence of the specific choice of the nonlinear interactions, probably aimed at reproducing atmospheric dynamics (e.g. Rossby waves \cite{MajdaWang2006}).
Specifically, the way the equations are built enforces the slow variables to produce waves propagating with phase velocity in the westward direction, and the fast ones in eastward direction.
We will see that response functions help to visualize in an alternative way the propagation of waves that are triggered by the initial perturbation. 

Fig.~\ref{fig:nondiag_resp_noneq} shows intra-scale energy response functions for slow (first row) and fast (second row) variables, when measured in the westward (left column) or eastward (right column) direction, with the usual rescaled times. 
The most stricking feature is the almost perfect similarity between the westward slow responses in panel (a) and eastward fast responses in panel (d): the functions show positive peaks that become more damped and delayed as one gets further away from the perturbation. 
As highlighted in the insets the speed of propagation is constant: the time to reach a site at a distance $d$ is linear in $d$. 
On the other hand, panels (b) and (c), showing respectively eastward slow responses and westward fast responses, show a less clear behaviour with both positive and negative peaks at short times, even though some qualitative similarities are observed. This coherent behaviour of the response functions in the direction of a statistical flux, accompanied by incoherence in the opposite direction, was also observed in turbulent shell models \cite{Cocciaglia2024nonequilibrium}. 
\begin{figure}[t]
    \includegraphics[width=0.5\textwidth]{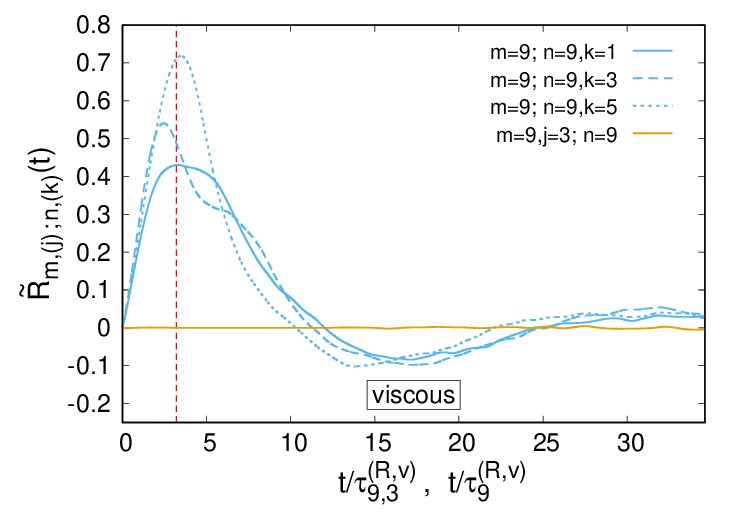}
    \vspace{-8mm}
    \caption{Rescaled inter-scale response functions \eqref{eq:rescaled_interscale_resp} in the viscous system. Azure plots show the effect of a slow-variable energy perturbation on three of its corresponding fast variables, the orange one the effect of a fast-variable perturbation on its corresponding slow variable. The time axis is rescaled with $\tau^{(R,v)}_{9,3}$ for the azure plots and with $\tau^{(R,v)}_{9}$ for the orange one. The red vertical dashed line corresponds to $t=\tau^{(R,v)}_{9}$.}
    \label{fig:resp_noneq_interscale}
\end{figure}
It is possible to provide an heuristic interpretation of the unclear behaviour in the ``upstream'' direction (opposite to that of wave propagation). For simplicity we focus on the responses of slow variables but the same reasoning also applies to fast ones.  
The excess of energy induced by a perturbation on a slow variable is transferred both to adjacent slow d.o.f.'s and to corresponding fast variables, where the fluctuations continue to propagate.
Thus, the contribution measured in the upstream slow responses comes both from transfers that took place at that scale, but also from fluctuations that have propagated downstream among the fast variables and then returned to the slow scales. 
This generates the apparently incoherent behaviour we observe in Figs.~\ref{fig:nondiag_resp_noneq}b and \ref{fig:nondiag_resp_noneq}c.
Let us remark, in passing, that the response functions closest to the perturbation display non-zero initial time derivatives. 
It is easy to see that these derivatives depend on third-order single-time correlation functions, which are positive on average due to the positively-skewed probability distribution of the variables. 
Thus, positive or negative initial slopes are determined by the signs of the nonlinear terms in Eq.~\eqref{eq:model}.

When studying inter-scale response functions some care must be taken, especially when the two scales are characterized by different amplitudes and timescales. 
In such cases a suitable rescaling is needed for comparing response functions with perturbation and measure on different variables. The new rescaled functions are:
\begin{equation}
    \tilde{R}_{m,(j)\hspace{.1em};\hspace{.1em} n,(k)}(t) = \frac{\delta E_{X_m (y_{m,j})}}{\langle E_{X_n (y_{n,k})} \rangle} R_{m,(j)\hspace{.1em};\hspace{.1em} n,(k)}(t) = \frac{\overline{\delta E_{X_n (y_{n,k})}}(t)}{\langle E_{X_n (y_{n,k})} \rangle}\ .
    \label{eq:rescaled_interscale_resp}
\end{equation}
They measure the relative energy deviation of a chosen variable caused by a perturbation on the energy of another (or the same) variable.
We show in Fig.~\ref{fig:resp_noneq_interscale} how the inter-scale response functions \eqref{eq:resp_Xy} and \eqref{eq:resp_yX} appear after the rescaling \eqref{eq:rescaled_interscale_resp}. 
It is rather clear that a finite perturbation on a fast variable has practically no effect on the corresponding slow one, while an energy fluctuation on a slow variable has a visible effect on the corresponding fast variables. 
Moreover, the maximum responses for the fast variables are attained at a time comparable with the relaxation time of the slow variable, $\tau^{(R,v)}_{9}$ (red vertical dashed line).
These results should make clear that, in order to give meaning to the measurement of a response function, one needs to know which degree of freedom has been initially perturbed, and possibly know the dynamical features (amplitude, timescale) of such variable. 

\begin{figure}[t]
    \includegraphics[width=0.5\textwidth]{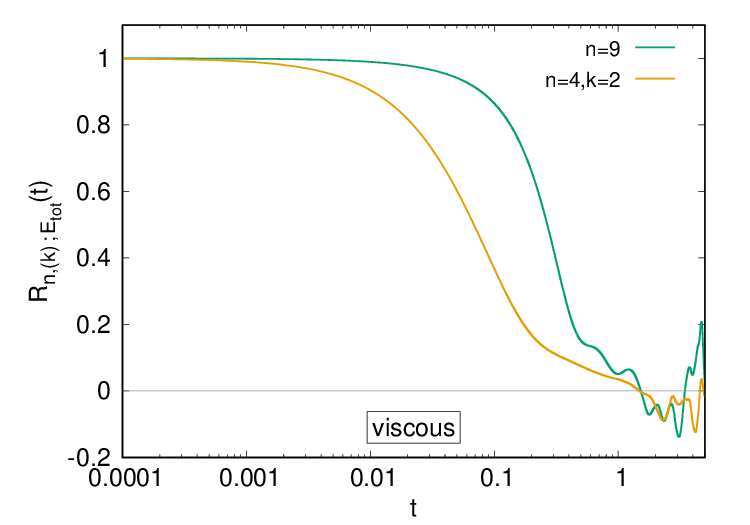}
    \vspace{-6mm}
    \caption{Response functions of the total energy $E_{tot}$ to two equal perturbations performed on $E_{X_9}$ (green curve) and $E_{y_{4,2}}$ (orange curve), displayed with logarithmic time axis. 
    }
    \label{fig:resp_noneq_enetot}
\end{figure}

In order to further prove this point, we show in Fig.~\ref{fig:resp_noneq_enetot} two response functions of the total energy $E_{tot}$ (defined in Eq.~\ref{eq:Energy_tot-Variation})
when a perturbation of same amplitude is performed on a slow or on a fast variable, namely:
\begin{equation}
    R_{m,(j)\hspace{.1em} ;\hspace{.1em} tot}(t) = \frac{\overline{\delta E_{tot}}(t)}{\delta E^{(m,(j))}}\,.
\end{equation}
The common perturbation has a strength: $\delta E^{(m,(j))} = \sqrt{\delta E_{X_m}\, \delta E_{y_{m,j}}}$, where $\delta E_{X_m}$ and $\delta E_{y_{m,j}}$ are the perturbations previously used respectively for slow and fast variables in Figs.~\ref{fig:nondiag_resp_noneq} and \ref{fig:resp_noneq_interscale}. 
The key observation here is that the two functions are subject to two perturbations of equal amplitudes (same denominator) and measure responses of the same quantity, yet they showcase very different relaxation behaviour: 
a faster relaxation is associated to the initial perturbation on the fast variable.
That is to say that measuring the induced fluctuations of a physical quantity cannot provide much information about the physics of the system if one does not know where the perturbation causing the fluctuations came from (i.e. which observable was perturbed). 

This is a fairly 
severe limitation when considering complex systems such as the climate one. For example, consider a situation in which one is interested in predicting how a fluctuation of the 2m-temperature (the temperature of the air measured $2$ meters above Earth's surface) on a tropical region evolves over time. Clearly the behaviour of this fluctuation is strongly influenced by its cause. If it was caused by an increase in sea surface temperature (as during ENSO phenomenon), then the time evolution of the fluctuation follows the evolution of this driving. If instead it was caused by an increase in the solar incident radiation, then its time scale would be linked to that of the radiative processes.

\section{Conclusion}\label{sec:conclusion}
In the present work we have studied in details two common indicators, namely asymmetric time-correlation functions and response functions, of time-reversal symmetry breaking in the Lorenz96 model, taken as a paradigm of evolution of geophysical flows. 
We have shown that higher order correlation functions, specifically third moments, successfully discriminate the statistical state of the system. 
In the inviscid case, the behaviour of both $H_{X_n^2,X_n^{}}(\tau)$ and $H_{y_{n,k}^2,y_{n,k}^{}}(\tau)$ is dominated by statistical errors causing modest fluctuations around zero as expected for an equilibrium system. 
In the scenario with forcing and dissipation, instead, these correlations show a pronounced peak at small times followed by damped oscillations, dispelling any doubts about the nonequilibrium nature of the dynamics. 
These correlation functions prove to be very powerful, since they allow to discriminate between equilibrium and nonequilibrium even without the knowledge of the evolution equations - which was not the case in the present study. One only needs a (stationary) time signal of the physical process, in terms of which one constructs suitable asymmetric time correlations.
We then focused on studying the responses to external perturbations. At variance with previous studies where spatiotemporally modulated perturbations are considered~\cite{Lucarini2008response,Lucarini2018revising}, we opted for impulsive (instantaneous) perturbations of local energies. 
It has been shown that, apart from providing useful information about the propagation phenomena in the system, the asymptotic behavior of the responses provides clear indications on the presence of temporal asymmetries. 
In this regard, since in the equilibrium system energy is a constant of motion, the excess energy due to the external perturbation is distributed among all degrees of freedom, such that, for long times, all the response functions converge to the same non-vanishing constant. 
In the viscous system, on the other hand, the initial excess of energy is progressively dampened until the energy of the system returns to fluctuate around the value of the unperturbed dynamics.  
As a fingerprint of this dissipation process,  in the limit $t\to\infty$ all response functions vanish. Regarding transport phenomena, the response functions clearly highlight the presence of traveling waves in both slow and fast variables propagating with constant speed but in opposite directions. The study of interscale responses then allowed us to elucidate some aspects of multiscale systems. 
In particular, due to the different order of magnitude of the variables involved in the viscous system, it has been observed that a finite perturbation of a fast variable induces negligible fluctuations (in percentage) in the slow variables, while perturbations of the slow variables cause macroscopically relevant fluctuations of the fast variables. 
Naturally any perturbation, however small, induces large discrepancies in the phase-space evolution due to the chaotic nature of the system.
Although the particular coupling between the slow and fast scales certainly plays a role, this characteristic is expected to be typical of many geophysical systems. 
Finally, by considering the response of global variables (such as the total energy of the system) to local perturbations of fixed amplitude we provided evidence of the impossibility of predicting the behaviour of the system without detailed knowledge of either its dynamics or the procedure adopted to produce the fluctuation. 
It is equally questionable to use results and features from time correlation functions, such as decorrelation times, to gain information about the relaxation following an external perturbation.
This consideration sheds light on some of the practical difficulties that must be faced when studying complex systems, such as climate, where it is impossible to directly perturb the system (except by considering computational models). 
Furthermore, in experimental records macroscopic fluctuations generally arise from the concomitance of different physical processes. 
An interesting perspective is to apply the tools investigated in this article to probe a more genuine geophysical system, if not directly experimental records. 
Indeed, higher order correlation functions offer important information about the statistics of the system that can be used in the context of stochastic parametrization. 
The response functions may be more difficult to obtain, but nevertheless they may provide useful insights about both transport and statistical properties of the system under investigation.


\begin{acknowledgments}
We wish to acknowledge M. Cencini and A. Vulpiani for useful suggestions and a careful reading of the manuscript.
\end{acknowledgments}


\section*{Author declarations}
\subsection*{Conflict of Interest}
The authors have no conflicts to disclose.

\subsection*{Author Contributions}
\noindent \textbf{Niccolò Cocciaglia}: Conceptualization (equal); Formal analysis (equal); Methodology (equal); Software (equal); Visualization (equal); Writing – original draft (equal).

\noindent \textbf{Dario Lucente}: Conceptualization (equal); Formal analysis (equal); Methodology (equal); Software (equal); Visualization (equal); Writing – original draft (equal).


\section*{Data Availability Statement}
The data that support the findings of this study are available from the corresponding author upon reasonable request.

\bibliography{bibliography}

\begin{thebibliography}{39}%
\makeatletter
\providecommand \@ifxundefined [1]{%
 \@ifx{#1\undefined}
}%
\providecommand \@ifnum [1]{%
 \ifnum #1\expandafter \@firstoftwo
 \else \expandafter \@secondoftwo
 \fi
}%
\providecommand \@ifx [1]{%
 \ifx #1\expandafter \@firstoftwo
 \else \expandafter \@secondoftwo
 \fi
}%
\providecommand \natexlab [1]{#1}%
\providecommand \enquote  [1]{``#1''}%
\providecommand \bibnamefont  [1]{#1}%
\providecommand \bibfnamefont [1]{#1}%
\providecommand \citenamefont [1]{#1}%
\providecommand \href@noop [0]{\@secondoftwo}%
\providecommand \href [0]{\begingroup \@sanitize@url \@href}%
\providecommand \@href[1]{\@@startlink{#1}\@@href}%
\providecommand \@@href[1]{\endgroup#1\@@endlink}%
\providecommand \@sanitize@url [0]{\catcode `\\12\catcode `\$12\catcode
  `\&12\catcode `\#12\catcode `\^12\catcode `\_12\catcode `\%12\relax}%
\providecommand \@@startlink[1]{}%
\providecommand \@@endlink[0]{}%
\providecommand \url  [0]{\begingroup\@sanitize@url \@url }%
\providecommand \@url [1]{\endgroup\@href {#1}{\urlprefix }}%
\providecommand \urlprefix  [0]{URL }%
\providecommand \Eprint [0]{\href }%
\providecommand \doibase [0]{http://dx.doi.org/}%
\providecommand \selectlanguage [0]{\@gobble}%
\providecommand \bibinfo  [0]{\@secondoftwo}%
\providecommand \bibfield  [0]{\@secondoftwo}%
\providecommand \translation [1]{[#1]}%
\providecommand \BibitemOpen [0]{}%
\providecommand \bibitemStop [0]{}%
\providecommand \bibitemNoStop [0]{.\EOS\space}%
\providecommand \EOS [0]{\spacefactor3000\relax}%
\providecommand \BibitemShut  [1]{\csname bibitem#1\endcsname}%
\let\auto@bib@innerbib\@empty
\bibitem [{\citenamefont {Lorenz}(2005)}]{Lorenz2005Designing}%
  \BibitemOpen
  \bibfield  {author} {\bibinfo {author} {\bibfnamefont {E.~N.}\ \bibnamefont
  {Lorenz}},\ }\bibfield  {title} {\enquote {\bibinfo {title} {Designing
  chaotic models},}\ }\href {\doibase 10.1175/JAS3430.1} {\bibfield  {journal}
  {\bibinfo  {journal} {J. Atmos. Sci.}\ }\textbf {\bibinfo {volume} {62}},\
  \bibinfo {pages} {1574 -- 1587} (\bibinfo {year} {2005})}\BibitemShut
  {NoStop}%
\bibitem [{\citenamefont {Lorenz}(1963)}]{Lorenz63Deterministic}%
  \BibitemOpen
  \bibfield  {author} {\bibinfo {author} {\bibfnamefont {E.~N.}\ \bibnamefont
  {Lorenz}},\ }\bibfield  {title} {\enquote {\bibinfo {title} {Deterministic
  nonperiodic flow},}\ }\href {\doibase
  10.1175/1520-0469(1963)020<0130:DNF>2.0.CO;2} {\bibfield  {journal} {\bibinfo
   {journal} {J.Atmos. Sci.}\ }\textbf {\bibinfo {volume} {20}},\ \bibinfo
  {pages} {130 -- 141} (\bibinfo {year} {1963})}\BibitemShut {NoStop}%
\bibitem [{\citenamefont {Lorenz}(1995)}]{Lorenz1996predictability}%
  \BibitemOpen
  \bibfield  {author} {\bibinfo {author} {\bibfnamefont {E.~N.}\ \bibnamefont
  {Lorenz}},\ }\bibfield  {title} {\enquote {\bibinfo {title} {Predictability:
  a problem partly solved},}\ }\href
  {https://www.ecmwf.int/en/elibrary/75462-predictability-problem-partly-solved}
  {\bibfield  {journal} {\bibinfo  {journal} {Seminar on Predictability, 4-8
  September 1995}\ }\textbf {\bibinfo {volume} {1}},\ \bibinfo {pages} {1--18}
  (\bibinfo {year} {1995})}\BibitemShut {NoStop}%
\bibitem [{\citenamefont {Boffetta}\ \emph {et~al.}(2000)\citenamefont
  {Boffetta}, \citenamefont {Celani}, \citenamefont {Cencini}, \citenamefont
  {Lacorata},\ and\ \citenamefont {Vulpiani}}]{boffetta2000predictability}%
  \BibitemOpen
  \bibfield  {author} {\bibinfo {author} {\bibfnamefont {G.}~\bibnamefont
  {Boffetta}}, \bibinfo {author} {\bibfnamefont {A.}~\bibnamefont {Celani}},
  \bibinfo {author} {\bibfnamefont {M.}~\bibnamefont {Cencini}}, \bibinfo
  {author} {\bibfnamefont {G.}~\bibnamefont {Lacorata}}, \ and\ \bibinfo
  {author} {\bibfnamefont {A.}~\bibnamefont {Vulpiani}},\ }\bibfield  {title}
  {\enquote {\bibinfo {title} {The predictability problem in systems with an
  uncertainty in the evolution law},}\ }\href {\doibase
  10.1088/0305-4470/33/7/302} {\bibfield  {journal} {\bibinfo  {journal} {J.
  Phys. A}\ }\textbf {\bibinfo {volume} {33}},\ \bibinfo {pages} {1313}
  (\bibinfo {year} {2000})}\BibitemShut {NoStop}%
\bibitem [{\citenamefont {Lacorata}\ and\ \citenamefont
  {Vulpiani}(2007)}]{Lacorata2007fluctuation}%
  \BibitemOpen
  \bibfield  {author} {\bibinfo {author} {\bibfnamefont {G.}~\bibnamefont
  {Lacorata}}\ and\ \bibinfo {author} {\bibfnamefont {A.}~\bibnamefont
  {Vulpiani}},\ }\bibfield  {title} {\enquote {\bibinfo {title}
  {Fluctuation-response relation and modeling in systems with fast and slow
  dynamics},}\ }\href {\doibase 10.5194/npg-14-681-2007} {\bibfield  {journal}
  {\bibinfo  {journal} {Nonlin. Processes Geophys.}\ }\textbf {\bibinfo
  {volume} {14}},\ \bibinfo {pages} {681--694} (\bibinfo {year}
  {2007})}\BibitemShut {NoStop}%
\bibitem [{\citenamefont {Karimi}\ and\ \citenamefont
  {Paul}(2010)}]{Karimi2012extensive}%
  \BibitemOpen
  \bibfield  {author} {\bibinfo {author} {\bibfnamefont {A.}~\bibnamefont
  {Karimi}}\ and\ \bibinfo {author} {\bibfnamefont {M.~R.}\ \bibnamefont
  {Paul}},\ }\bibfield  {title} {\enquote {\bibinfo {title} {{Extensive chaos
  in the Lorenz-96 model}},}\ }\href {\doibase 10.1063/1.3496397} {\bibfield
  {journal} {\bibinfo  {journal} {Chaos}\ }\textbf {\bibinfo {volume} {20}},\
  \bibinfo {pages} {043105} (\bibinfo {year} {2010})}\BibitemShut {NoStop}%
\bibitem [{\citenamefont {Gallavotti}\ and\ \citenamefont
  {Lucarini}(2014)}]{Gallavotti2014equivalence}%
  \BibitemOpen
  \bibfield  {author} {\bibinfo {author} {\bibfnamefont {G.}~\bibnamefont
  {Gallavotti}}\ and\ \bibinfo {author} {\bibfnamefont {V.}~\bibnamefont
  {Lucarini}},\ }\bibfield  {title} {\enquote {\bibinfo {title} {Equivalence of
  non-equilibrium ensembles and representation of friction in turbulent flows:
  {T}he {L}orenz 96 model},}\ }\href {\doibase 10.1007/s10955-014-1051-6}
  {\bibfield  {journal} {\bibinfo  {journal} {J. Stat. Phys.}\ }\textbf
  {\bibinfo {volume} {156}},\ \bibinfo {pages} {1027--1065} (\bibinfo {year}
  {2014})}\BibitemShut {NoStop}%
\bibitem [{\citenamefont {van Kekem}\ and\ \citenamefont
  {Sterk}(2018)}]{vankekem2018wave}%
  \BibitemOpen
  \bibfield  {author} {\bibinfo {author} {\bibfnamefont {D.~L.}\ \bibnamefont
  {van Kekem}}\ and\ \bibinfo {author} {\bibfnamefont {A.~E.}\ \bibnamefont
  {Sterk}},\ }\bibfield  {title} {\enquote {\bibinfo {title} {Wave propagation
  in the {L}orenz-96 model},}\ }\href {\doibase 10.5194/npg-25-301-2018}
  {\bibfield  {journal} {\bibinfo  {journal} {Nonlin. Processes Geophys.}\
  }\textbf {\bibinfo {volume} {25}},\ \bibinfo {pages} {301--314} (\bibinfo
  {year} {2018})}\BibitemShut {NoStop}%
\bibitem [{\citenamefont {Orrell}\ and\ \citenamefont
  {Smith}(2003)}]{Orrell2003}%
  \BibitemOpen
  \bibfield  {author} {\bibinfo {author} {\bibfnamefont {D.}~\bibnamefont
  {Orrell}}\ and\ \bibinfo {author} {\bibfnamefont {L.~A.}\ \bibnamefont
  {Smith}},\ }\bibfield  {title} {\enquote {\bibinfo {title} {Visualising
  bifurcations in high dimensional systems: {T}he spectral bifurcation
  diagram},}\ }\href {\doibase 10.1142/S0218127403008387} {\bibfield  {journal}
  {\bibinfo  {journal} {Int. J. Bifurcat. Chaos}\ }\textbf {\bibinfo {volume}
  {13}},\ \bibinfo {pages} {3015--3027} (\bibinfo {year} {2003})}\BibitemShut
  {NoStop}%
\bibitem [{\citenamefont {Stappers}\ and\ \citenamefont
  {Barkmeijer}(2012)}]{Stappers2012}%
  \BibitemOpen
  \bibfield  {author} {\bibinfo {author} {\bibfnamefont {R.~J.~J.}\
  \bibnamefont {Stappers}}\ and\ \bibinfo {author} {\bibfnamefont
  {J.}~\bibnamefont {Barkmeijer}},\ }\bibfield  {title} {\enquote {\bibinfo
  {title} {Optimal linearization trajectories for tangent linear models},}\
  }\href {\doibase https://doi.org/10.1002/qj.908} {\bibfield  {journal}
  {\bibinfo  {journal} {Q. J. Roy. Meteor. Soc.}\ }\textbf {\bibinfo {volume}
  {138}},\ \bibinfo {pages} {170--184} (\bibinfo {year} {2012})}\BibitemShut
  {NoStop}%
\bibitem [{\citenamefont {de~Leeuw}\ \emph {et~al.}(2018)\citenamefont
  {de~Leeuw}, \citenamefont {Dubinkina}, \citenamefont {Frank}, \citenamefont
  {Steyer}, \citenamefont {Tu},\ and\ \citenamefont
  {Van~Vleck}}]{deLeeuw2018projected}%
  \BibitemOpen
  \bibfield  {author} {\bibinfo {author} {\bibfnamefont {B.}~\bibnamefont
  {de~Leeuw}}, \bibinfo {author} {\bibfnamefont {S.}~\bibnamefont {Dubinkina}},
  \bibinfo {author} {\bibfnamefont {J.}~\bibnamefont {Frank}}, \bibinfo
  {author} {\bibfnamefont {A.}~\bibnamefont {Steyer}}, \bibinfo {author}
  {\bibfnamefont {X.}~\bibnamefont {Tu}}, \ and\ \bibinfo {author}
  {\bibfnamefont {E.}~\bibnamefont {Van~Vleck}},\ }\bibfield  {title} {\enquote
  {\bibinfo {title} {Projected shadowing-based data assimilation},}\ }\href
  {\doibase 10.1137/17M1141163} {\bibfield  {journal} {\bibinfo  {journal}
  {SIAM J. Appl. Dyn. Syst.}\ }\textbf {\bibinfo {volume} {17}},\ \bibinfo
  {pages} {2446--2477} (\bibinfo {year} {2018})}\BibitemShut {NoStop}%
\bibitem [{\citenamefont {Kerin}\ and\ \citenamefont
  {Engler}(2020)}]{kerin2020lorenz}%
  \BibitemOpen
  \bibfield  {author} {\bibinfo {author} {\bibfnamefont {J.}~\bibnamefont
  {Kerin}}\ and\ \bibinfo {author} {\bibfnamefont {H.}~\bibnamefont {Engler}},\
  }\href@noop {} {\enquote {\bibinfo {title} {On the {L}orenz '96 model and
  some generalizations},}\ } (\bibinfo {year} {2020}),\ \Eprint
  {http://arxiv.org/abs/2005.07767} {arXiv:2005.07767 [math.DS]} \BibitemShut
  {NoStop}%
\bibitem [{\citenamefont {Basnarkov}\ and\ \citenamefont
  {Kocarev}(2012)}]{Basnarkov2012forecast}%
  \BibitemOpen
  \bibfield  {author} {\bibinfo {author} {\bibfnamefont {L.}~\bibnamefont
  {Basnarkov}}\ and\ \bibinfo {author} {\bibfnamefont {L.}~\bibnamefont
  {Kocarev}},\ }\bibfield  {title} {\enquote {\bibinfo {title} {Forecast
  improvement in {L}orenz 96 system},}\ }\href {\doibase
  10.5194/npg-19-569-2012} {\bibfield  {journal} {\bibinfo  {journal} {Nonlin.
  Processes Geophys.}\ }\textbf {\bibinfo {volume} {19}},\ \bibinfo {pages}
  {569--575} (\bibinfo {year} {2012})}\BibitemShut {NoStop}%
\bibitem [{\citenamefont {Sterk}\ \emph {et~al.}(2012)\citenamefont {Sterk},
  \citenamefont {Holland}, \citenamefont {Rabassa}, \citenamefont {Broer},\
  and\ \citenamefont {Vitolo}}]{Sterk2012predictability}%
  \BibitemOpen
  \bibfield  {author} {\bibinfo {author} {\bibfnamefont {A.~E.}\ \bibnamefont
  {Sterk}}, \bibinfo {author} {\bibfnamefont {M.~P.}\ \bibnamefont {Holland}},
  \bibinfo {author} {\bibfnamefont {P.}~\bibnamefont {Rabassa}}, \bibinfo
  {author} {\bibfnamefont {H.~W.}\ \bibnamefont {Broer}}, \ and\ \bibinfo
  {author} {\bibfnamefont {R.}~\bibnamefont {Vitolo}},\ }\bibfield  {title}
  {\enquote {\bibinfo {title} {Predictability of extreme values in geophysical
  models},}\ }\href {\doibase 10.5194/npg-19-529-2012} {\bibfield  {journal}
  {\bibinfo  {journal} {Nonlin. Processes Geophys.}\ }\textbf {\bibinfo
  {volume} {19}},\ \bibinfo {pages} {529--539} (\bibinfo {year}
  {2012})}\BibitemShut {NoStop}%
\bibitem [{\citenamefont {Carlu}\ \emph {et~al.}(2019)\citenamefont {Carlu},
  \citenamefont {Ginelli}, \citenamefont {Lucarini},\ and\ \citenamefont
  {Politi}}]{carlu2019lyapunov}%
  \BibitemOpen
  \bibfield  {author} {\bibinfo {author} {\bibfnamefont {M.}~\bibnamefont
  {Carlu}}, \bibinfo {author} {\bibfnamefont {F.}~\bibnamefont {Ginelli}},
  \bibinfo {author} {\bibfnamefont {V.}~\bibnamefont {Lucarini}}, \ and\
  \bibinfo {author} {\bibfnamefont {A.}~\bibnamefont {Politi}},\ }\bibfield
  {title} {\enquote {\bibinfo {title} {Lyapunov analysis of multiscale
  dynamics: the slow bundle of the two-scale {L}orenz 96 model},}\ }\href
  {\doibase 10.5194/npg-26-73-2019} {\bibfield  {journal} {\bibinfo  {journal}
  {Nonlin. Processes Geophys.}\ }\textbf {\bibinfo {volume} {26}},\ \bibinfo
  {pages} {73--89} (\bibinfo {year} {2019})}\BibitemShut {NoStop}%
\bibitem [{\citenamefont {Pomeau}(1982)}]{Pomeau1982symetrie}%
  \BibitemOpen
  \bibfield  {author} {\bibinfo {author} {\bibfnamefont {Y.}~\bibnamefont
  {Pomeau}},\ }\bibfield  {title} {\enquote {\bibinfo {title} {Sym\'etrie des
  fluctuations dans le renversement du temps},}\ }\href {\doibase
  10.1051/jphys:01982004306085900} {\bibfield  {journal} {\bibinfo  {journal}
  {J. Phys. France}\ }\textbf {\bibinfo {volume} {43}},\ \bibinfo {pages}
  {859--867} (\bibinfo {year} {1982})}\BibitemShut {NoStop}%
\bibitem [{\citenamefont {Kubo}(1966)}]{kubo1966FDT}%
  \BibitemOpen
  \bibfield  {author} {\bibinfo {author} {\bibfnamefont {R.}~\bibnamefont
  {Kubo}},\ }\bibfield  {title} {\enquote {\bibinfo {title} {The
  fluctuation-dissipation theorem},}\ }\href {\doibase
  10.1088/0034-4885/29/1/306} {\bibfield  {journal} {\bibinfo  {journal} {Rep.
  Prog. Phys.}\ }\textbf {\bibinfo {volume} {29}},\ \bibinfo {pages} {255}
  (\bibinfo {year} {1966})}\BibitemShut {NoStop}%
\bibitem [{\citenamefont {Marconi}\ \emph {et~al.}(2008)\citenamefont
  {Marconi}, \citenamefont {Puglisi}, \citenamefont {Rondoni},\ and\
  \citenamefont {Vulpiani}}]{FDreport}%
  \BibitemOpen
  \bibfield  {author} {\bibinfo {author} {\bibfnamefont {U.~M.~B.}\
  \bibnamefont {Marconi}}, \bibinfo {author} {\bibfnamefont {A.}~\bibnamefont
  {Puglisi}}, \bibinfo {author} {\bibfnamefont {L.}~\bibnamefont {Rondoni}}, \
  and\ \bibinfo {author} {\bibfnamefont {A.}~\bibnamefont {Vulpiani}},\
  }\bibfield  {title} {\enquote {\bibinfo {title} {Fluctuation–dissipation:
  Response theory in statistical physics},}\ }\href {\doibase
  https://doi.org/10.1016/j.physrep.2008.02.002} {\bibfield  {journal}
  {\bibinfo  {journal} {Phys. Rep.}\ }\textbf {\bibinfo {volume} {461}},\
  \bibinfo {pages} {111--195} (\bibinfo {year} {2008})}\BibitemShut {NoStop}%
\bibitem [{\citenamefont {Lucente}\ \emph
  {et~al.}(2023{\natexlab{a}})\citenamefont {Lucente}, \citenamefont {Viale},
  \citenamefont {Gnoli}, \citenamefont {Puglisi},\ and\ \citenamefont
  {Vulpiani}}]{Lucente2023revealing}%
  \BibitemOpen
  \bibfield  {author} {\bibinfo {author} {\bibfnamefont {D.}~\bibnamefont
  {Lucente}}, \bibinfo {author} {\bibfnamefont {M.}~\bibnamefont {Viale}},
  \bibinfo {author} {\bibfnamefont {A.}~\bibnamefont {Gnoli}}, \bibinfo
  {author} {\bibfnamefont {A.}~\bibnamefont {Puglisi}}, \ and\ \bibinfo
  {author} {\bibfnamefont {A.}~\bibnamefont {Vulpiani}},\ }\bibfield  {title}
  {\enquote {\bibinfo {title} {Revealing the nonequilibrium nature of a
  granular intruder: The crucial role of non-gaussian behavior},}\ }\href
  {\doibase 10.1103/PhysRevLett.131.078201} {\bibfield  {journal} {\bibinfo
  {journal} {Phys. Rev. Lett.}\ }\textbf {\bibinfo {volume} {131}},\ \bibinfo
  {pages} {078201} (\bibinfo {year} {2023}{\natexlab{a}})}\BibitemShut
  {NoStop}%
\bibitem [{\citenamefont {Cocciaglia}, \citenamefont {Cencini},\ and\
  \citenamefont {Vulpiani}(2024)}]{Cocciaglia2024nonequilibrium}%
  \BibitemOpen
  \bibfield  {author} {\bibinfo {author} {\bibfnamefont {N.}~\bibnamefont
  {Cocciaglia}}, \bibinfo {author} {\bibfnamefont {M.}~\bibnamefont {Cencini}},
  \ and\ \bibinfo {author} {\bibfnamefont {A.}~\bibnamefont {Vulpiani}},\
  }\bibfield  {title} {\enquote {\bibinfo {title} {Nonequilibrium statistical
  mechanics of the turbulent energy cascade: Irreversibility and response
  functions},}\ }\href {\doibase 10.1103/PhysRevE.109.014113} {\bibfield
  {journal} {\bibinfo  {journal} {Phys. Rev. E}\ }\textbf {\bibinfo {volume}
  {109}},\ \bibinfo {pages} {014113} (\bibinfo {year} {2024})}\BibitemShut
  {NoStop}%
\bibitem [{\citenamefont {Frank}\ \emph {et~al.}(2014)\citenamefont {Frank},
  \citenamefont {Mitchell}, \citenamefont {Dodds},\ and\ \citenamefont
  {Danforth}}]{frank2014standing}%
  \BibitemOpen
  \bibfield  {author} {\bibinfo {author} {\bibfnamefont {M.~R.}\ \bibnamefont
  {Frank}}, \bibinfo {author} {\bibfnamefont {L.}~\bibnamefont {Mitchell}},
  \bibinfo {author} {\bibfnamefont {P.~S.}\ \bibnamefont {Dodds}}, \ and\
  \bibinfo {author} {\bibfnamefont {C.~M.}\ \bibnamefont {Danforth}},\
  }\bibfield  {title} {\enquote {\bibinfo {title} {Standing swells surveyed
  showing surprisingly stable solutions for the {L}orenz '96 model},}\ }\href
  {\doibase 10.1142/S0218127414300274} {\bibfield  {journal} {\bibinfo
  {journal} {Int. J. Bifurc. Chaos}\ }\textbf {\bibinfo {volume} {24}},\
  \bibinfo {pages} {1430027} (\bibinfo {year} {2014})}\BibitemShut {NoStop}%
\bibitem [{\citenamefont {Bohr}\ \emph {et~al.}(1998)\citenamefont {Bohr},
  \citenamefont {Jensen}, \citenamefont {Paladin},\ and\ \citenamefont
  {Vulpiani}}]{bohr1998dynamical}%
  \BibitemOpen
  \bibfield  {author} {\bibinfo {author} {\bibfnamefont {T.}~\bibnamefont
  {Bohr}}, \bibinfo {author} {\bibfnamefont {M.~H.}\ \bibnamefont {Jensen}},
  \bibinfo {author} {\bibfnamefont {G.}~\bibnamefont {Paladin}}, \ and\
  \bibinfo {author} {\bibfnamefont {A.}~\bibnamefont {Vulpiani}},\ }\href@noop
  {} {\emph {\bibinfo {title} {Dynamical Systems Approach to Turbulence}}}\
  (\bibinfo  {publisher} {Cambridge University Press, New York},\ \bibinfo
  {year} {1998})\BibitemShut {NoStop}%
\bibitem [{\citenamefont {Biferale}(2003)}]{biferale2003shell}%
  \BibitemOpen
  \bibfield  {author} {\bibinfo {author} {\bibfnamefont {L.}~\bibnamefont
  {Biferale}},\ }\bibfield  {title} {\enquote {\bibinfo {title} {Shell models
  of energy cascade in turbulence},}\ }\href {\doibase
  https://doi.org/10.1146/annurev.fluid.35.101101.161122} {\bibfield  {journal}
  {\bibinfo  {journal} {Annual review of fluid mechanics}\ }\textbf {\bibinfo
  {volume} {35}},\ \bibinfo {pages} {441--468} (\bibinfo {year}
  {2003})}\BibitemShut {NoStop}%
\bibitem [{\citenamefont {Ditlevsen}(2010)}]{ditlevsen2010turbulence}%
  \BibitemOpen
  \bibfield  {author} {\bibinfo {author} {\bibfnamefont {P.~D.}\ \bibnamefont
  {Ditlevsen}},\ }\href@noop {} {\emph {\bibinfo {title} {Turbulence and shell
  models}}}\ (\bibinfo  {publisher} {Cambridge University Press},\ \bibinfo
  {year} {2010})\BibitemShut {NoStop}%
\bibitem [{\citenamefont {Lebowitz}(1999)}]{lebowitz1999microscopic}%
  \BibitemOpen
  \bibfield  {author} {\bibinfo {author} {\bibfnamefont {J.~L.}\ \bibnamefont
  {Lebowitz}},\ }\bibfield  {title} {\enquote {\bibinfo {title} {Microscopic
  origins of irreversible macroscopic behavior},}\ }\href {\doibase
  https://doi.org/10.1016/S0378-4371(98)00514-7} {\bibfield  {journal}
  {\bibinfo  {journal} {Physica A}\ }\textbf {\bibinfo {volume} {263}},\
  \bibinfo {pages} {516--527} (\bibinfo {year} {1999})}\BibitemShut {NoStop}%
\bibitem [{\citenamefont {Lebowitz}\ and\ \citenamefont
  {Spohn}(1999)}]{lebowitz1999gallavotti}%
  \BibitemOpen
  \bibfield  {author} {\bibinfo {author} {\bibfnamefont {J.~L.}\ \bibnamefont
  {Lebowitz}}\ and\ \bibinfo {author} {\bibfnamefont {H.}~\bibnamefont
  {Spohn}},\ }\bibfield  {title} {\enquote {\bibinfo {title} {A
  {G}allavotti--{C}ohen-type symmetry in the large deviation functional for
  stochastic dynamics},}\ }\href {\doibase 10.1023/A:1004589714161} {\bibfield
  {journal} {\bibinfo  {journal} {J. Stat. Phys.}\ }\textbf {\bibinfo {volume}
  {95}},\ \bibinfo {pages} {333--365} (\bibinfo {year} {1999})}\BibitemShut
  {NoStop}%
\bibitem [{\citenamefont {Lucente}\ \emph
  {et~al.}(2023{\natexlab{b}})\citenamefont {Lucente}, \citenamefont {Puglisi},
  \citenamefont {Viale},\ and\ \citenamefont {Vulpiani}}]{Lucente2023poisson}%
  \BibitemOpen
  \bibfield  {author} {\bibinfo {author} {\bibfnamefont {D.}~\bibnamefont
  {Lucente}}, \bibinfo {author} {\bibfnamefont {A.}~\bibnamefont {Puglisi}},
  \bibinfo {author} {\bibfnamefont {M.}~\bibnamefont {Viale}}, \ and\ \bibinfo
  {author} {\bibfnamefont {A.}~\bibnamefont {Vulpiani}},\ }\bibfield  {title}
  {\enquote {\bibinfo {title} {Statistical features of systems driven by
  non-gaussian processes: theory \& practice},}\ }\href {\doibase
  10.1088/1742-5468/ad063b} {\bibfield  {journal} {\bibinfo  {journal} {J.
  Stat. Mech.}\ }\textbf {\bibinfo {volume} {2023}},\ \bibinfo {pages} {113202}
  (\bibinfo {year} {2023}{\natexlab{b}})}\BibitemShut {NoStop}%
\bibitem [{\citenamefont {Josserand}\ \emph {et~al.}(2017)\citenamefont
  {Josserand}, \citenamefont {Le~Berre}, \citenamefont {Lehner},\ and\
  \citenamefont {Pomeau}}]{Josserand2017}%
  \BibitemOpen
  \bibfield  {author} {\bibinfo {author} {\bibfnamefont {C.}~\bibnamefont
  {Josserand}}, \bibinfo {author} {\bibfnamefont {M.}~\bibnamefont {Le~Berre}},
  \bibinfo {author} {\bibfnamefont {T.}~\bibnamefont {Lehner}}, \ and\ \bibinfo
  {author} {\bibfnamefont {Y.}~\bibnamefont {Pomeau}},\ }\bibfield  {title}
  {\enquote {\bibinfo {title} {Turbulence: Does energy cascade exist?}}\ }\href
  {\doibase 10.1007/s10955-016-1642-5} {\bibfield  {journal} {\bibinfo
  {journal} {J. Stat. Phys.}\ }\textbf {\bibinfo {volume} {167}},\ \bibinfo
  {pages} {596--625} (\bibinfo {year} {2017})}\BibitemShut {NoStop}%
\bibitem [{\citenamefont {Ramsey}\ and\ \citenamefont
  {Rothman}(1996)}]{Ramsey1996business}%
  \BibitemOpen
  \bibfield  {author} {\bibinfo {author} {\bibfnamefont {J.~B.}\ \bibnamefont
  {Ramsey}}\ and\ \bibinfo {author} {\bibfnamefont {P.}~\bibnamefont
  {Rothman}},\ }\bibfield  {title} {\enquote {\bibinfo {title} {Time
  irreversibility and business cycle asymmetry},}\ }\href
  {http://www.jstor.org/stable/2077963} {\bibfield  {journal} {\bibinfo
  {journal} {J. Money Credit Banking}\ }\textbf {\bibinfo {volume} {28}},\
  \bibinfo {pages} {1--21} (\bibinfo {year} {1996})}\BibitemShut {NoStop}%
\bibitem [{\citenamefont {Kubo}, \citenamefont {Toda},\ and\ \citenamefont
  {Hashitsume}(2012)}]{kubo2012statistical}%
  \BibitemOpen
  \bibfield  {author} {\bibinfo {author} {\bibfnamefont {R.}~\bibnamefont
  {Kubo}}, \bibinfo {author} {\bibfnamefont {M.}~\bibnamefont {Toda}}, \ and\
  \bibinfo {author} {\bibfnamefont {N.}~\bibnamefont {Hashitsume}},\
  }\href@noop {} {\emph {\bibinfo {title} {Statistical physics II:
  nonequilibrium statistical mechanics}}},\ Vol.~\bibinfo {volume} {31}\
  (\bibinfo  {publisher} {Springer Science \& Business Media},\ \bibinfo {year}
  {2012})\BibitemShut {NoStop}%
\bibitem [{\citenamefont {Lorenz}\ and\ \citenamefont
  {Emanuel}(1998)}]{Lorenz1998optimal}%
  \BibitemOpen
  \bibfield  {author} {\bibinfo {author} {\bibfnamefont {E.~N.}\ \bibnamefont
  {Lorenz}}\ and\ \bibinfo {author} {\bibfnamefont {K.~A.}\ \bibnamefont
  {Emanuel}},\ }\bibfield  {title} {\enquote {\bibinfo {title} {Optimal sites
  for supplementary weather observations: Simulation with a small model},}\
  }\href {\doibase 10.1175/1520-0469(1998)055<0399:OSFSWO>2.0.CO;2} {\bibfield
  {journal} {\bibinfo  {journal} {J. Atmos. Sci.}\ }\textbf {\bibinfo {volume}
  {55}},\ \bibinfo {pages} {399 -- 414} (\bibinfo {year} {1998})}\BibitemShut
  {NoStop}%
\bibitem [{\citenamefont {Majda}\ and\ \citenamefont
  {Wang}(2006)}]{MajdaWang2006}%
  \BibitemOpen
  \bibfield  {author} {\bibinfo {author} {\bibfnamefont {A.}~\bibnamefont
  {Majda}}\ and\ \bibinfo {author} {\bibfnamefont {X.}~\bibnamefont {Wang}},\
  }\enquote {\bibinfo {title} {Nonlinear dynamics and statistical theories for
  basic geophysical flows},}\ \ (\bibinfo  {publisher} {Cambridge University
  Press},\ \bibinfo {year} {2006})\ Chap.\ \bibinfo {chapter} {7.4}\BibitemShut
  {NoStop}%
\bibitem [{\citenamefont {Baldovin}, \citenamefont {Cecconi},\ and\
  \citenamefont {Vulpiani}(2020)}]{Baldovin2020understanding}%
  \BibitemOpen
  \bibfield  {author} {\bibinfo {author} {\bibfnamefont {M.}~\bibnamefont
  {Baldovin}}, \bibinfo {author} {\bibfnamefont {F.}~\bibnamefont {Cecconi}}, \
  and\ \bibinfo {author} {\bibfnamefont {A.}~\bibnamefont {Vulpiani}},\
  }\bibfield  {title} {\enquote {\bibinfo {title} {Understanding causation via
  correlations and linear response theory},}\ }\href {\doibase
  10.1103/PhysRevResearch.2.043436} {\bibfield  {journal} {\bibinfo  {journal}
  {Phys. Rev. Res.}\ }\textbf {\bibinfo {volume} {2}},\ \bibinfo {pages}
  {043436} (\bibinfo {year} {2020})}\BibitemShut {NoStop}%
\bibitem [{\citenamefont {Sarra}, \citenamefont {Baldovin},\ and\ \citenamefont
  {Vulpiani}(2021)}]{Sarra2021response}%
  \BibitemOpen
  \bibfield  {author} {\bibinfo {author} {\bibfnamefont {C.}~\bibnamefont
  {Sarra}}, \bibinfo {author} {\bibfnamefont {M.}~\bibnamefont {Baldovin}}, \
  and\ \bibinfo {author} {\bibfnamefont {A.}~\bibnamefont {Vulpiani}},\
  }\bibfield  {title} {\enquote {\bibinfo {title} {Response and flux of
  information in extended nonequilibrium dynamics},}\ }\href {\doibase
  10.1103/PhysRevE.104.024116} {\bibfield  {journal} {\bibinfo  {journal}
  {Phys. Rev. E}\ }\textbf {\bibinfo {volume} {104}},\ \bibinfo {pages}
  {024116} (\bibinfo {year} {2021})}\BibitemShut {NoStop}%
\bibitem [{\citenamefont {Ruelle}(1998)}]{Ruelle1998general}%
  \BibitemOpen
  \bibfield  {author} {\bibinfo {author} {\bibfnamefont {D.}~\bibnamefont
  {Ruelle}},\ }\bibfield  {title} {\enquote {\bibinfo {title} {General linear
  response formula in statistical mechanics, and the fluctuation-dissipation
  theorem far from equilibrium},}\ }\href {\doibase
  https://doi.org/10.1016/S0375-9601(98)00419-8} {\bibfield  {journal}
  {\bibinfo  {journal} {Phys. Lett. A}\ }\textbf {\bibinfo {volume} {245}},\
  \bibinfo {pages} {220--224} (\bibinfo {year} {1998})}\BibitemShut {NoStop}%
\bibitem [{\citenamefont {Lucarini}(2008)}]{Lucarini2008response}%
  \BibitemOpen
  \bibfield  {author} {\bibinfo {author} {\bibfnamefont {V.}~\bibnamefont
  {Lucarini}},\ }\bibfield  {title} {\enquote {\bibinfo {title} {Response
  theory for equilibrium and non-equilibrium statistical mechanics: Causality
  and generalized {K}ramers-{K}ronig relations},}\ }\href {\doibase
  10.1007/s10955-008-9498-y} {\bibfield  {journal} {\bibinfo  {journal} {J.
  Stat. Phys.}\ }\textbf {\bibinfo {volume} {131}},\ \bibinfo {pages}
  {543--558} (\bibinfo {year} {2008})}\BibitemShut {NoStop}%
\bibitem [{\citenamefont {Lucarini}\ and\ \citenamefont
  {Sarno}(2011)}]{Lucarini2011statistical}%
  \BibitemOpen
  \bibfield  {author} {\bibinfo {author} {\bibfnamefont {V.}~\bibnamefont
  {Lucarini}}\ and\ \bibinfo {author} {\bibfnamefont {S.}~\bibnamefont
  {Sarno}},\ }\bibfield  {title} {\enquote {\bibinfo {title} {A statistical
  mechanical approach for the computation of the climatic response to general
  forcings},}\ }\href {\doibase 10.5194/npg-18-7-2011} {\bibfield  {journal}
  {\bibinfo  {journal} {Nonlin. Processes Geophys.}\ }\textbf {\bibinfo
  {volume} {18}},\ \bibinfo {pages} {7--28} (\bibinfo {year}
  {2011})}\BibitemShut {NoStop}%
\bibitem [{\citenamefont {Lucarini}(2018)}]{Lucarini2018revising}%
  \BibitemOpen
  \bibfield  {author} {\bibinfo {author} {\bibfnamefont {V.}~\bibnamefont
  {Lucarini}},\ }\bibfield  {title} {\enquote {\bibinfo {title} {Revising and
  extending the linear response theory for statistical mechanical systems:
  Evaluating observables as predictors and predictands},}\ }\href {\doibase
  10.1007/s10955-018-2151-5} {\bibfield  {journal} {\bibinfo  {journal} {J.
  Stat. Phys.}\ }\textbf {\bibinfo {volume} {173}},\ \bibinfo {pages}
  {1698–1721} (\bibinfo {year} {2018})}\BibitemShut {NoStop}%
\bibitem [{\citenamefont {Boffetta}\ \emph {et~al.}(2003)\citenamefont
  {Boffetta}, \citenamefont {Lacorata}, \citenamefont {Musacchio},\ and\
  \citenamefont {Vulpiani}}]{Rfinpert}%
  \BibitemOpen
  \bibfield  {author} {\bibinfo {author} {\bibfnamefont {G.}~\bibnamefont
  {Boffetta}}, \bibinfo {author} {\bibfnamefont {G.}~\bibnamefont {Lacorata}},
  \bibinfo {author} {\bibfnamefont {S.}~\bibnamefont {Musacchio}}, \ and\
  \bibinfo {author} {\bibfnamefont {A.}~\bibnamefont {Vulpiani}},\ }\bibfield
  {title} {\enquote {\bibinfo {title} {Relaxation of finite perturbations:
  Beyond the fluctuation-response relation},}\ }\href {\doibase
  10.1063/1.1579643} {\bibfield  {journal} {\bibinfo  {journal} {Chaos}\
  }\textbf {\bibinfo {volume} {13}},\ \bibinfo {pages} {806--811} (\bibinfo
  {year} {2003})}\BibitemShut {NoStop}%
\end{thebibliography}%

\end{document}